\documentstyle[emulateapj,epsf]{article}
\lefthead{Balogh et al.}
\righthead{Recent Star Formation in Cluster and Field Galaxies at $z\approx0.3$}

\begin{document}
\newcommand{\oii}{[OII]$\lambda$3727}
\newcommand{\ow}{$W_{\circ}(OII)$}
\newcommand{\hd}{$W_{\circ}(H\delta)$}
\newcommand{\balm}{$W_{\circ}(H\delta\gamma)$}
\newcommand{\hdlo}{$W_{\circ}(H\delta)_{low}$}
\newcommand{\hdhi}{$W_{\circ}(H\delta)_{high}$}
\newcommand{\hdem}{$W_{\circ}(H\delta)_{em}$}
\newcommand{\hdabs}{$W_{\circ}(H\delta)_{abs}$}
\newcommand{\balmlo}{$W_{\circ}(H\delta\gamma)_{low}$}
\newcommand{\balmhi}{$W_{\circ}(H\delta\gamma)_{high}$}
\newcommand{\hg}{$W_{\circ}(H\gamma)$}
\newcommand{\ebv}{$E_{\footnotesize \bv}$}
\newcommand{\br}{{\em B-R}}
\submitted{ApJ, Accepted June 28, 1999}
\title{Differential Galaxy Evolution in Cluster and Field Galaxies at $z\approx0.3$}

\author{Michael L. Balogh\altaffilmark{1},
Simon L. Morris\altaffilmark{2, 5},}
\author{H. K. C. Yee\altaffilmark{3, 5},
R.G. Carlberg\altaffilmark{3, 5}, 
and Erica Ellingson\altaffilmark{4, 5}}
\altaffiltext{1}{\small{Department of Physics \& Astronomy, University of Victoria, Victoria, BC, V8X 4M6, Canada. \\ email: balogh@uvphys.phys.uvic.ca}}

\altaffiltext{2}{\small{Dominion Astrophysical Observatory, National Research Council, 5071 West Saanich Road, Victoria, B.C., V8X 4M6 Canada. email: Simon.Morris@hia.nrc.ca}}

\altaffiltext{3}{\small{Department of Astronomy, University of Toronto, Toronto, Ontario, M5S 1A7 Canada. \\ email: hyee, carlberg@astro.utoronto.ca}}

\altaffiltext{4}{\small{CASA, University of Colorado, Boulder, Colorado 80309-0389. \\ email: e.elling@casa.colorado.edu}}
\altaffiltext{5}{\small{Visiting Astronomer, Canada--France--Hawaii Telescope, which is operated by the National Research Council of Canada, le Centre Nationale de la Recherche Scientifique, and the University of Hawaii.}}

\begin{abstract}
We measure spectral indices for
1823 galaxies in the CNOC1 sample of 
fifteen X--ray luminous clusters at $0.18<z<0.55$, to 
investigate the mechanisms responsible for differential evolution
between the galaxy cluster and field environments. 
The radial trends of D4000, \hd\ and \ow\ are all consistent with an age sequence,
in the sense that the last episode of star formation occurred more recently in 
galaxies farthest from the cluster center.  Throughout the cluster environment, galaxies
show evidence for older stellar populations than field galaxies; they have weaker \ow\ and
\hd\ lines, and stronger D4000 indices.

From our primary sample of 1413 galaxies, statistically corrected for incompleteness and selection
effects, we identify a sample of K+A galaxies,
which have strong H$\delta$ absorption lines (\hd\ $>5$ \AA) but no [\ion{O}{2}] emission (\ow\ $<5$ \AA), 
perhaps indicative of recently terminated star formation.
The observed fraction of $4.4\pm0.7$\% in the cluster sample is an overestimate
due to a systematic effect which results from the large uncertainties on individual 
spectral index measurements.
Corrected for this bias, we estimate that K+A galaxies make up only 2.1$\pm$0.7\% of 
 the cluster sample, and $0.1\pm0.7$\% of
the field.    From the subsample of galaxies more luminous than $M_r=-18.8 + 5\log{h}$, 
which is statistically representative of a complete sample to this limit, the corrected fraction of
K+A galaxies is 1.5$\pm$0.8\% in the cluster, and 1.2$\pm$0.8\% in the field.
Compared with
the $z\approx0.1$ fraction of 0.30\% (\cite{Z+96}), the fraction of K+A galaxies in the CNOC1 field sample
is greater by perhaps a factor of four, but with only 1$\sigma$
significance; no further evolution of this fraction is detectable over our redshift range.  

We compare our data with the results of PEGASE and GISSEL96 spectrophotometric
models and conclude, from the relative fractions of red and blue galaxies with
no \oii\ emission and strong H$\delta$ absorption, that up to $1.9\pm0.8$\%
of the cluster population may have had its star formation recently truncated without a starburst.
However, this is still not significantly greater than the fraction of such galaxies in the field, 3.1$\pm1.0$\%.
Furthermore, we do not detect an excess of cluster galaxies that have unambiguously
undergone a starburst within the last 1 Gyr.  In fact, at 6.3$\pm2.1$\%, the A+em galaxies that
Poggianti et al. (1999) have recently suggested are dusty starbursts are twice as common
in the field as in the cluster environment. 

Our results imply
that these cluster environments are not responsible for inducing starbursts;
thus, the increase in cluster blue galaxy fraction with redshift may not be a strictly cluster--specific phenomenon.
We suggest that the truncation of star formation in clusters may largely be a gradual process, perhaps 
due to the exhaustion of 
gas in the galactic disk over fairly long timescales; in this case differential evolution may result because
field galaxies can refuel their disk with gas from an extended halo, thus regenerating
star formation, while cluster galaxies may not have such a halo and so continue to
evolve passively.
\end{abstract}


\section{Introduction} \label{sec-intro}
The formation and evolution histories of the galaxies that populate rare, rich clusters
are currently unknown; however, it seems unlikely that a single history can accurately
describe the whole population (e.g., \cite{H+98}).  The giant elliptical galaxies which
dominate the central regions of rich clusters seem to have been in place for well
over 5 Gyr, and have only evolved passively since (e.g., \cite{Ellis97,BKT,Barger98}).  The remainder of the population
is dominated in numbers by dwarf spheroidal galaxies and in mass by early type
galaxies (E, S0, early spirals).  There is strong theoretical (e.g., \cite{GG,AC94,VH93,T98}) and observational
(e.g., \cite{ZF93,HJ96}) evidence that much of the cluster galaxy population has been built up by the
accretion of galaxies from the surrounding, low density, field environment, which implies that
the rich cluster environment has affected a morphological change in these galaxies.
A clear correlation between galaxy morphology with local density (\cite{Dressler,D+97}) and
cluster--centric radius (\cite{WGJ}) has been observed.  The well--known lack
of emission line galaxies in clusters (\cite{O60,G78,DTS}) has also been shown to correlate with cluster--centric
radius (Balogh et al. 1997) and to be not entirely accounted for by the morphology--radius
relation (\cite{MW,B+98}).  Recently, Abraham et al. (1996)\nocite{A2390} and Morris et al. (1998)\nocite{1621} 
showed that these
various effects could be interpreted as an ``age'' gradient, in the sense that galaxies
farther from the cluster center formed part of their stellar component more recently than
the central galaxies.  

Another key piece of evidence that cluster galaxies are evolving strongly is
the observation that the fraction of blue, spiral galaxies in clusters increases
with redshift (e.g., \cite{BO78}, 1984, \cite{DG83,D+94,C+94,RS95,Y+95,C+98}, but see \cite{S+98}).  This is usually referred
to as the Butcher--Oemler (B--O) effect.  To avoid possible misunderstanding, 
we will always refer to the B--O effect as this
exclusively, a separate effect from the above problem of differential evolution between the
cluster and field, though the two phenomena may be related.  It
has been shown that the star formation rate in the field also increases
with redshift (\cite{BES88,CFRSVI,RR+97,TM98,CNOC2}); thus, the B--O effect as we consider it
here may not be an exclusively cluster--specific effect and
cannot na$\ddot{\mbox{\i}}$vely be used as an explanation of the different populations which inhabit
cluster and field environments.  Furthermore, it is not clear that the B--O effect can be directly
interpreted as an evolutionary effect; Kauffmann (1995)\nocite{K95} has shown that rich 
clusters at $z\approx0.4$ are not the direct progenitors of low redshift, comparably rich clusters.
More recently, Andreon \& Ettori (1999)\nocite{AE99} have shown
that the higher redshift Butcher--Oemler clusters have higher X--ray luminosities than their
low redshift counterparts.  Since fair samples of clusters generally show either no evolution, or even {\it negative}
evolution
in the X--ray luminosity function (e.g., \cite{Henry92,Collins97,V+98,RDN98}), this confirms
 that the high redshift Butcher--Oemler 
clusters may not be the evolutionary predecessors of the low redshift clusters.

Various mechanisms may be responsible for the transformation of a star--forming field galaxy
to a passively evolving cluster member.  An infalling
galaxy which passes through or near the center of the cluster may have much of
its gas content stripped by the hot intra--cluster medium (ICM), following which
star formation will cease (\cite{N82,BD86,BLO}).  If only gas in an extended,
diffuse halo is stripped, star formation may be allowed to continue by consuming the remaining
disk gas but, without infall to replenish this supply, star formation will die out on timescales of
a few Gyr (\cite{LTC}).  Interaction with the ICM near the virial radius,
where the ICM density is not high enough to completely strip the galaxy of its gas, may
induce intense bursts of star formation (\cite{DG83,E91,GJ87}) which use up the remaining gas
supply.  Such bursts may also be induced by  interactions with nearby galaxies (\cite{M+96},1998, \cite{F98})
or, near the cluster center, by interactions with the tidal field of the cluster (\cite{BV90,V93,HB96,F98}).

Understanding the mechanisms responsible for this change in star 
formation history is necessary to appropriately correct the CNOC1 universal mass
density estimate (\cite{CNOC1}) for the difference in mass--to--light (M/L) ratio
in clusters, compared with the field.  The correction applied in Carlberg, Yee
\& Ellingson (1997) could be compromised if a large number of cluster galaxies underwent
strong starbursts which do {\it not} occur in the field population.  

Dressler \& Gunn (1983) identified a galaxy population which they later termed E+A
galaxies, that have strong Balmer lines but no detectable emission lines.  These 
galaxies (which we will hereafter term K+A to be consistent with current nomenclature)
appear to have recently terminated star formation, within the $\sim$1 Gyr prior to observation.
As such, they play an important part in understanding the evolutionary history of galaxy
populations.  Evidence for these (and related) galaxies have been found both locally
(\cite{C+93,Z+96}) and at moderate redshift (e.g., \cite{S+85,LH86,CS87,MEC,BES88,FMB91,A2390,F+98,1621,D+99}), in both cluster and field environments.
Many authors have interpreted the presence of these galaxies as evidence that a large
percentage of cluster galaxies recently underwent strong starbursts (e.g., \cite{DG83,CS87,Barger,PB96,C+98,P+99}).  
However, it is still debatable whether or not starburst and post--starburst galaxies 
observed in clusters are unique to these environments.  Furthermore, other authors have
claimed that most K+A galaxies can largely be explained as the result
of truncated star formation, without a large, initial burst (\cite{NBK,A2390,1621}).
It is necessary both to distinguish between these two
scenarios, and to compare their relative importance in the cluster and field environments,
to determine which physical processes are plausible explanations for the differential
evolution of cluster and field galaxies, and to determine how these processes
will affect the M/L ratio in these environments.

The CNOC1 survey  (Yee, Ellingson \& Carlberg 1996) provides an excellent sample of galaxy spectra for which a reliable estimate
of the star formation history for clusters at $z\approx0.3$ can be determined.  This
is a result of (1) a large sample size (over 2000 spectra); (2) well understood, and 
properly correctable, selection
effects; (3) an identically selected field galaxy sample over
the same redshift range; (4) objective measurements of spectral indices with well understood
and empirically calibrated uncertainties; and (5) consistent comparison between data and spectrophotometric
models.  From these data, we will attempt to specifically address the following questions.
First,  what fraction of galaxies in the CNOC1 sample, regardless of environment, had significant star formation truncated
in the last billion years or so?  If this is truly larger than the low redshift
fraction, it suggests a link with the B--O effect.  Secondly, how does the abundance
of these galaxies in the field  compare with the abundance in clusters?  An overabundance
in clusters would suggest that the mechanism responsible for generating these 
galaxies may be what leads to the differential evolution of cluster and field galaxies.
Finally, {\em how} is the star formation in these galaxies truncated?
We will use spectrophotometric models to try to
distinguish between a) truncation following a short burst; or, b) abrupt
truncation without a burst.  

The paper is organized as follows.  In \S\ref{sec-data} we define our data sample
and spectral index measurements.  In particular, we discuss the details of our sample 
selection, statistical corrections, magnitude limits, and index uncertainties.  Interpretation of our line indices
is made based on a comparison with PEGASE (\cite{PEGASE}) spectrophotometric models,
described in \S\ref{sec-models}.  Our galaxy classification scheme is presented in \S\ref{sec-defs},
and is based on the line indices of local galaxies and results
of the PEGASE models (detailed in the Appendix).
We present our results in various forms in \S\ref{sec-res}, and take care to explore systematic
effects which result from the large uncertainties on our line indices.  In \S\ref{sec-compare}
we compare our results with the work of Zabludoff et al. (1996), Barger et al. (1996), Dressler et al. (1999) and others,
and show that there is some consistency of results, though considerable variety in interpretation.
We discuss some of the implications of these results in \S\ref{sec-discuss},
and summarize our findings in \S\ref{sec-conc}.

\section{Observations and Measurements} \label{sec-data}
\subsection{A Review of the Full CNOC1 Sample}\label{sec-cnocsample}
The CNOC1 cluster sample consists of CFHT MOS images (Gunn $g$ and $r$) and
spectra of galaxies in the fields of fifteen\footnote{Omitting 
cluster E0906+11, for which a velocity dispersion could not be computed due to an apparent double component structure(\cite{CNOC1}).}
 X--ray luminous clusters in the redshift
range $0.18<z<0.55$, selected from the EMSS survey\footnote{Except Abell 2390, which is not in the
EMSS sample, but of comparable richness and X--ray luminosity to the other CNOC1 clusters (\cite{A2390}).} (\cite{GL94}).  
The observational strategy and full details of the survey can be found in Yee et al. (1996).
In particular, Yee et al. carefully consider and detail the selection effects 
of this sample, and describe how these are corrected for.  For
convenience, we repeat here some of the details in that paper relevant to the
present analysis.

The spectra are obtained with the O300 grism in place, which results in a dispersion of 
3.45\AA\ per pixel and covers the rest--frame wavelength range from (approximately)
\oii\ to the G--band ($\lambda\approx4300$\AA) in most galaxies.  Slits are 1\farcs5 wide, which
results in a spectral resolution of about 16.5\AA\ FWHM.  Images and spectra of galaxies
in each cluster are obtained from either one, three or five MOS fields mosaiced
east--west or north--south, which provide non--uniform
coverage as far out as 1--2 $R_{200}$ in projected distance for most clusters,
where $R_{200}$ is the radius ($\approx1.2$ h$^{-1}$ Mpc) at which the mean interior mass density is equal to
200 times the critical density, and within which it is expected that the galaxies are
in virial equilibrium (\cite{GG}; \cite{CER}).  For $\Omega_\circ=0.2$, the cluster virial
radius is equal to approximately 1.2$R_{200}$.  The final catalog contains redshifts
for about 2500 galaxies, including field galaxies located in front of and behind each cluster.

\begin{figure*}
\begin{center}
\leavevmode \epsfysize=8cm \epsfbox{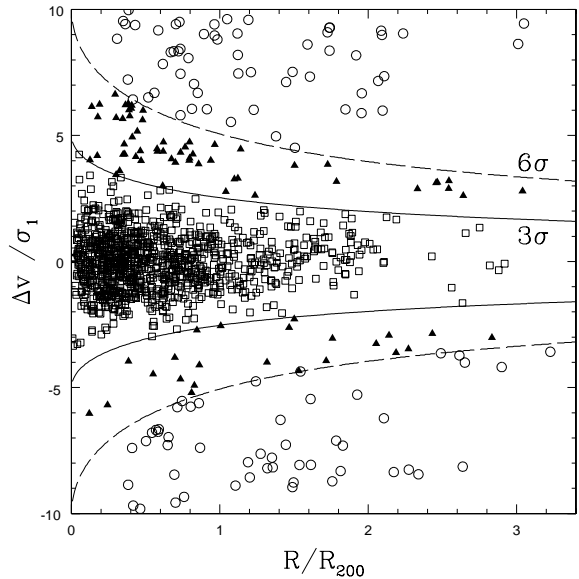}
\end{center}
\caption{This figure shows how we define our cluster and field samples.  The data plotted are
radial velocity differences from the BCG, normalized to the cluster velocity dispersion, $\sigma_1$,
against projected radii from the BCG, in units of $R_{200}$.  The {\it solid} and {\it dashed lines}
are the 3$\sigma(r)$ and 6$\sigma(r)$ contours, respectively, of the cluster mass model
of Carlberg et al. (1997).  We define cluster members ({\it squares}) as those which lie
within 3$\sigma(r)$, and field galaxies ({\it circles}) as those with velocities greater
than 6$\sigma(r)$.  The intermediate population ({\it solid triangles}) are termed ``near--field''.
The extent of the vertical axis is limited to clearly show the cluster population; only a
small fraction of the field sample is shown in this figure.
\label{fig-velrrr}}
\end{figure*}

In order to average over effects of non--sphericity, we will combine all fifteen clusters
to produce a single sample, hereafter referred to as {\it the cluster}.  
Cluster membership is determined based on the observed radial velocity difference
from the  brightest cluster galaxy (BCG\footnote{Except for cluster MS 0451.5+0250, for which no
redshift is available for the BCG.  The velocities and positions are measured relative to the
mean for this cluster.}).  We use the mass model of Carlberg et al. (1997)
to determine the (projected) radial dependence of the cluster velocity dispersion, $\sigma(r)$,
from the average, measured dispersion, $\sigma_1$.
In Figure \ref{fig-velrrr} we plot the galaxy velocities, normalized to $\sigma_1$, 
against their projected radius from the BCG, 
normalized to $R_{200}$.  The $3\sigma$ and $6\sigma$ contours of the mass model
are shown as the solid and dashed lines, respectively.  Galaxies with normalized
velocities less than  3$\sigma(r)$ are considered cluster members (squares), while those
with normalized velocities greater than 6$\sigma(r)$ comprise our field sample (circles).
The population with intermediate velocities 
is represented by the solid triangles, and are considered ``near--field''.  The 
extent of the vertical axis in this figure is limited for clarity; hence only a small
fraction of the total field population is displayed.  The cluster population appears
well separated from the field in this figure, and exhibits a rising velocity
dispersion toward the cluster center.  We will only include galaxies within 2$R_{200}$
in our cluster sample, as larger radii are much more sparsely sampled.

\begin{figure*}
\begin{center}
\leavevmode \epsfysize=8cm \epsfbox{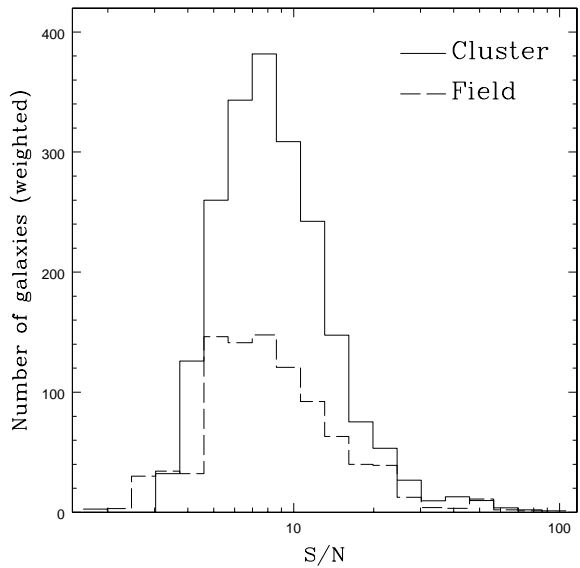}
\end{center}
\caption{The distribution of signal--to--noise per pixel is shown for the cluster {\it (solid line)} and field
{\it (dashed line)} samples.  The number of galaxies in each bin is weighted as described in \S\ref{sec-weights} and,
thus, represents the number of galaxies in the photometric sample, though the distribution is computed from
the spectroscopic sample.
\label{fig-snr}}
\end{figure*}

By extracting the spectra with standard IRAF\footnote{IRAF is distributed by the National Optical Astronomy Observatories which is 
operated by AURA 
Inc. under contract with NSF.} routines, an error vector is created
which includes Poisson errors, read noise
and sky subtraction uncertainties.  This allows a good determination of both the signal--to--noise (S/N)
ratio of the data, and also reliable errors on any measured quantities (see \S\ref{sec-indices}).
The S/N ratio is defined as the mean value of the flux per pixel divided by the error per pixel,
in the wavelength range $4050<\lambda/$\AA$<4250$.  
The resulting S/N distributions for the cluster and field samples are shown in
Figure \ref{fig-snr}.  The mode of the
distribution is at S/N$\approx$7 per pixel (or 15 per resolution element), 
and 94\% of spectra have S/N$\gtrsim3$.

The entire CNOC1 data sample, including the raw data and all measured quantities
included in this paper, will soon be available from the Canadian Astronomy Data
Centre (http://cadcwww.hia.nrc.ca/).  All cosmologically dependent calculations in this
paper are made assuming $q_\circ=0.1$, $\Lambda=0$.  The Hubble constant
is parameterized as usual by $h=H_\circ/100$ km s$^{-1}$ Mpc$^{-1}$.

\subsection{Spectral Index Definitions}\label{sec-indices}
We will consider three spectral indices, tabulated in Table \ref{tab-indices}, which are 
automatically measured by an IRAF {\em IMFORT}
routine available from the first author on request. 
The break strength at 4000\AA\ (D4000) is defined as the ratio of the flux in the
red continuum to that in the blue continuum.  We have chosen a much narrower definition
of this break than the standard definition (\cite{Hamilton}).  There are two principal
reasons for this.  First, the uncertainty in the standard index value, assuming Poisson noise,
was found to be a poor representation of the difference between multiple measurements
of the same galaxy (this analysis is discussed further in \S\ref{sec-error}).  
Reducing the width of the continuum definitions significantly improved
this correspondence.  Secondly, 
the new index is much less sensitive to reddening effects, which may be important (see
\S\ref{sec-variations}).

\begin{deluxetable}{cccc}
\tablewidth{0pt}
\footnotesize
\tablecaption{\label{tab-indices}}
\tablehead{\colhead{Index}&\colhead{Blue continuum}&\colhead{Line}&\colhead{Red continuum}\nl
           \colhead{}&\colhead{(\AA)}&\colhead{(\AA)}&\colhead{(\AA)}}
\startdata
 D4000   &  3850--3950 & --- & 4000--4100\nl
 [OII]$\lambda$3727 &  3653--3713 & 3713--3741 & 3741--3801\nl
 $H\delta$  &  4030--4082 & 4082--4122 & 4122--4170\nl
\enddata
\tablecomments{Definitions of line indices measured on the data and spectrophotometric
models.  The D4000 index is the ratio of the flux in the red continuum to that in the
blue continuum.  The other three indices are equivalent widths of the line over the 
wavelength range listed in Column 3, where the continuum level is estimated from the flux
blueward (Column 2) and redward (Column 4) of the line.}
\end{deluxetable}

There are considerable 
advantages in using D4000 instead of the available {\em g-r} color.  Most importantly, computing
a rest frame color from the observed photometry requires calculating K--corrections from
the very models we will be using to determine star formation histories and, thus, introduces
a circular argument.  Furthermore, the galaxy
spectra are obtained from slits which only sample the central regions of the galaxy, while the 
photometry represents the integrated light from the whole galaxy.  Thus, the photometric
colors and observed spectral features may not arise from the same environments. (However, at $z=0.3$, the slit
width of $1\farcs5$ corresponds to a linear size of 4.4$h^{-1}$ kpc, which is a sizable region).
In addition,
especially for our unusually narrow definition of D4000, this index is very reddening
insensitive.  The principal
disadvantage is that the measurement uncertainties in the D4000 measurements are generally larger than
those of the {\em g-r} colors.

The rest frame equivalent widths of the $H\delta$ Balmer 
absorption line, \hd, and the \oii\ emission line, \ow,
were automatically computed by summing the observed flux (accounting for partial pixels)
above (\ow) or below (\hd) the continuum level, which itself is estimated by fitting a straight line
to the flux in the continuum regions.  The continuum data are weighted by the inverse square of
the noise (from the IRAF noise vector), and the fit is then constructed by weighted linear regression.     
The blue and red continuum regions are defined in 
Table \ref{tab-indices}; the line flux is summed over the region defined ``Line'' in the
same table.   {\em Note that the \ow\ index is positive when the line is in emission, while the
\hd\ index is positive when the line is in absorption.}  

It must be noted that the indices defined above (or, indeed, any such indices) are not necessarily
related simply or directly to physical quantities. 
The value of the \hd\ index is very sensitive to the line and continuum
definitions, and also on the spectral resolution and sampling.  In particular, negative values of \hd\ do not always 
indicate emission, since absorption lines in the continuum regions (especially in galaxies
with old stellar populations) may
result in negative \hd\ index values.  Thus, caution must be taken when
comparing these measurements with those of other authors; in fact, the index definition
here differs slightly from that used by our own collaboration 
in Abraham et al. (1996).  

\subsubsection{Index Uncertainties}\label{sec-error}
Errors on the spectral indices are determined from the noise vector generated from the optimal
IRAF extraction.  The D4000 index is simply a flux ratio and, hence, the error is 
determined from standard propagation techniques.  Errors on equivalent width measurements
are computed from equation A8 in Bohlin et al. (1983); this makes use of the uncertainties
of the weighted, linear continuum fit, and the uncertainties of each pixel defined as
comprising the line itself.
Often, spectral regions contaminated by bright night sky line emission are very noisy, or 
interpolated over by hand.  Since the computed error vector reaches very large values at the wavelengths
of night sky lines, this effect will be reflected in the index uncertainties.  Excluding from
the sample all galaxies in which either of the lines at 5577 \AA, 5890 \AA, or 6300 \AA\ lands
within the index definitions of  $H\delta$ or [\ion{O}{2}] does not significantly change our
main results.  

\begin{figure*}
\begin{center}
\leavevmode \epsfysize=8cm \epsfbox{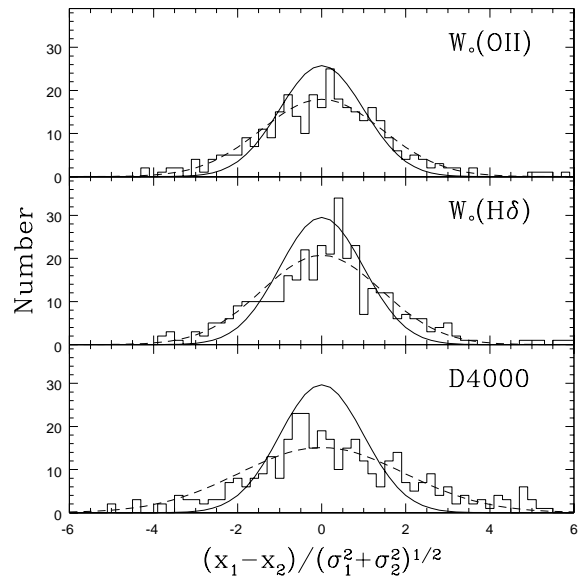}
\end{center}
\caption{For multiply observed galaxies in the sample, we plot the ratio of the difference
between two independent measurements of each index and the quadrature sum of the estimated index errors.
If the true errors are normally distributed and well represented by the estimated errors,
the distribution will be Gaussian with a mean of zero and variance of unity, shown by the
{\it solid line}.  The distributions are wider than this Gaussian, but consistent with one
of variance 1.44 (\ow), 1.42 (\hd) and 1.96 (D4000), shown as the {\it dashed lines}.
\label{fig-ind_err}}
\end{figure*}

Since spectra were obtained with up to three masks for each cluster and, also, since
adjacent fields overlap by about 15\arcsec, some galaxies were deliberately observed more than
once.  This allows us to test the reliability of our error estimate, by comparing the
difference between two measurements ($x_1-x_2$) with the quadrature sum of their errors,
$\sqrt{\sigma_1^2+\sigma_2^2}$.  
In Figure \ref{fig-ind_err} we plot the distribution of the ratio of these two numbers, 
$\epsilon=(x_1-x_2)/\sqrt{\sigma_1^2+\sigma_2^2}$, for each
of the three indices.  If the true errors are Gaussian distributed with a variance given by our error
estimate, the distribution of $\epsilon$ should be  Gaussian with a mean of zero and variance of unity; this
is shown as the solid curve for reference.  All of the $\epsilon$ distributions have a variance greater than
1, and are inconsistent with the solid curves.  This implies that our raw uncertainties are underestimated, which
may reflect the small but inevitable systematic error resulting from the subtraction of incorrect sky
levels in some cases.
The 3$\sigma$ clipped variance of each $\epsilon$ distribution is 1.44 for \ow, 1.42 for \hd\, and 1.96 for D4000.  
The distributions are better represented by a Gaussian of the corresponding, wider variance, shown as the dashed lines.
To compensate for this effect, we  multiplied all error estimates by the appropriate value.
The final index value adopted for multiply observed galaxies is the average, weighted by the square
of the error, of all independent measurements (up to three).

\subsection{The Data Sample}\label{sec-sample}
In order to carry out our analysis, statistical weights must be computed for all galaxies, to correct
for the incompleteness of the spectroscopic sampling; this procedure is reviewed in \S\ref{sec-weights}.  
We then proceed to define a luminosity--limited
sample, in \S\ref{sec-maglim}, which will be considered for those analyses which are sensitive to such
a limit.  For the remainder of the analysis, for which
the primary purpose is to compare properties in the cluster and field samples, we define a larger sample
which is not luminosity limited, but in which the lowest quality data are excluded, in \S\ref{sec-maxsamp}.
Unless otherwise specified, this latter is the data sample that will be discussed in the remainder of the paper.

\subsubsection{Data Weights}\label{sec-weights}
Since spectra are not obtained for all of the galaxies observed photometrically, we must
correct for this incompleteness and any selection effects that may arise as a result.
These effects, and the calculation of compensating statistical weights, are considered in detail in Yee et al. (1996);
we review them briefly here.
The main selection criterion is apparent magnitude; a smaller fraction of faint galaxies are observed
spectroscopically, relative to brighter galaxies.  The magnitude weight $W_m$ compensates for this effect.
A second order, geometric weight $W_{xy}$ is computed which compensates for such effects as the undersampling
of denser regions, and vignetting near the corners of the chip.  Finally, a color weight $W_c$ is computed
to account for the fact that bluer galaxies are more likely to show emission lines, facilitating
redshift determination; however, this is a small effect, and we have checked that the inclusion of this
weight does not significantly affect any of the results discussed in the present paper.  Both $W_{xy}$ and
$W_{c}$ are normalized so that their mean is 1.0 for the full sample.
In all of our statistical analysis, each galaxy in our sample is weighted by $W_{\rm spec}=W_m\times W_{xy}\times W_c$ to 
statistically correct for these selection effects\footnote{All uncertainties in this paper that are based
on $\sqrt{N}$ Poisson statistics are computed from the unweighted number of galaxies under consideration.}.
We have also computed a ``ring'' weight,
which is meant to correct for the fact that, due to the geometry of the cluster mosaics,
the outer regions of clusters are less well sampled (in area) than the central regions.  This correction is only
important in the presence of radial gradients, if a globally averaged cluster quantity is sought.  We will not
use this weight in the present analysis, 
though we have verified that its use does not change any of the results (recall 
that we restrict the cluster sample to those galaxies within $R<2R_{200}$).

\subsubsection{Luminosity Limited Sample}\label{sec-maglim}
\begin{figure*}
\begin{center}
\leavevmode \epsfysize=8cm \epsfbox{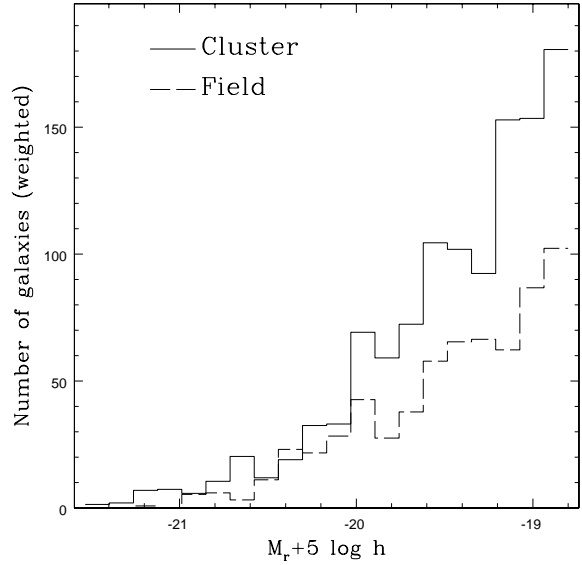}
\end{center}
\caption{The evolution and k--corrected, absolute Gunn r magnitude distributions of the 
cluster ({\it solid line}) and field ({\it dashed line}), luminosity limited galaxy samples.  Each
bin is weighted by $W_{\rm spec}$, discussed in \S\ref{sec-weights}; thus, the height of each
bin represents the number of galaxies in the photometric sample, as determined from the
spectroscopic sample.
\label{fig-Mag_maglim}}
\end{figure*}

We will first define a sample limited only in luminosity.  
Absolute $r$ magnitudes ($M_r$) are calculated from the photometry, and k--corrections
are made based on the {\em g-r} colors and the model spectral energy distributions of Coleman, Wu \& Weedman (1980),
convolved with the filter response function, for four,
non--evolving spectral types (E/S0, Sbc, Scd and Im).  We also correct $M_r$ for redshift 
evolution by assuming galaxies brighten by a factor (1+z) (\cite{L+98}; \cite{A+98}); thus, we correct all
luminosities to the corresponding $z=0$ values by dividing them by (1+z), though
this does not significantly affect our results.  
We chose an absolute magnitude limit of $M_r=-18.8+5 \log{h}$, since this excludes most of the
galaxies with large magnitude weights, $W_m>5$; this limit corresponds to 1.8 mag below
M$^*$ in $r$ (\cite{KE85}).
The absolute
magnitude distribution of the cluster sample (weighted by $W_{\rm spec}$) is compared with that of the field in
Figure \ref{fig-Mag_maglim}.  There are 1125 galaxies in this sample (omitting the BCGs and cluster members
beyond 2$R_{200}$), including 710 cluster members,
343 field galaxies and 69 near field galaxies.

\begin{figure*}
\begin{center}
\leavevmode \epsfysize=8cm \epsfbox{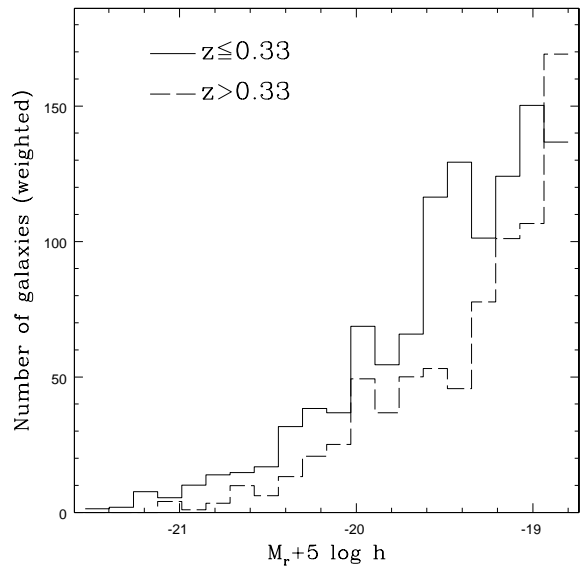}
\end{center}
\caption{The evolution and k--corrected, absolute Gunn r magnitude distributions of the 
low redshift ({\it solid line}) and high redshift ({\it dashed line}), luminosity limited galaxy 
samples.   Due both to the luminosity evolution correction, and to the fact that longer
exposure times were used for the higher
redshift clusters, the two distributions are similar.  The histograms are weighted by $W_{\rm spec}$;
see the caption of Figure \ref{fig-Mag_maglim}.
\label{fig-magz}}
\end{figure*}

The absolute magnitude distributions of the low ($z\leq 0.33$)
and high ($z>0.33$) redshift galaxies in this sample are shown in Figure \ref{fig-magz}.  As a result 
of the fainter apparent magnitude limit in the high redshift cluster samples (\cite{YEC}),
and the luminosity evolution correction, the two distributions are comparable.

\subsubsection{The Maximal Sample}\label{sec-maxsamp}
\begin{figure*}
\begin{center}
\leavevmode \epsfysize=8cm \epsfbox{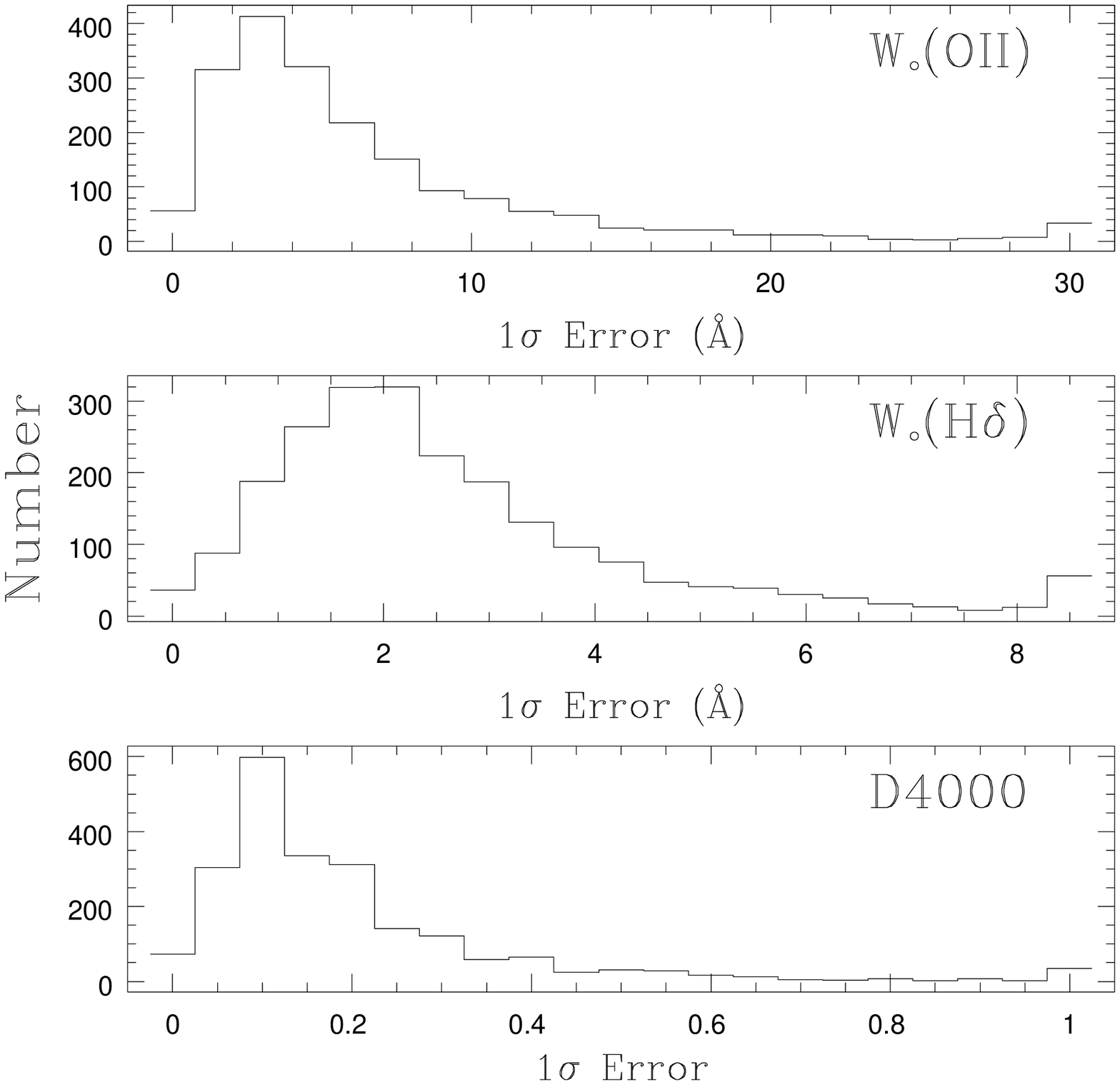}
\end{center}
\caption{The distribution of 1$\sigma$ errors on each of the three indices considered.
In each case, the tail of the distribution extends to errors much greater than the
mean.  To avoid including data with very large uncertainties, we exclude the $\sim10$\% of galaxies 
with errors greater than 15\AA, 5\AA, and 0.5\AA\ for \ow, \hd\ and D4000,
respectively, in the maximal sample.  This selection criteria is compensated for by
a statistical weight, $W_{\rm err}$, discussed in \S\ref{sec-maxsamp}.  
\label{fig-errordist}}
\end{figure*}

\begin{figure*}
\begin{center}
\leavevmode \epsfysize=8cm \epsfbox{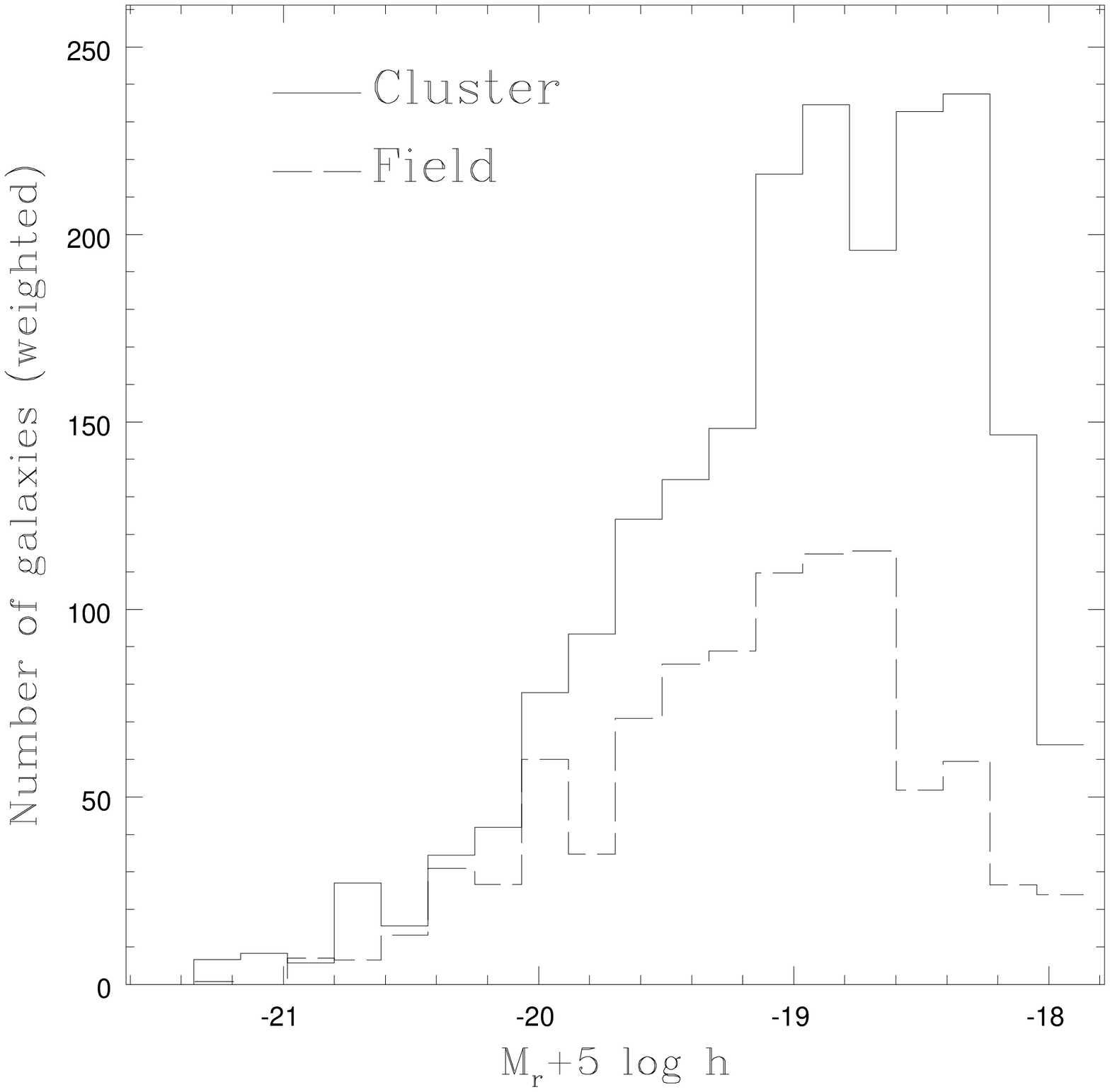}
\end{center}
\caption{The evolution and k--corrected, absolute Gunn r magnitude distributions of the 
cluster ({\it solid line}) and field ({\it dashed line}) maximal galaxy samples (\S\ref{sec-maxsamp}).  
The histograms are weighted by $W_{\rm spec}\times W_{\rm err}$, as discussed in
the text.  The
cluster sample is biased to lower luminosities relative to the field sample,
and a standard K--S test rejects the hypothesis that
the two are drawn from the same population with $\sim$3$\sigma$ confidence.
\label{fig-Mag_cf}}
\end{figure*}

For the purposes of comparing cluster and field galaxy populations, it is desirable to use as much
of the data as possible, excluding only that of the poorest quality, and ensuring that the luminosity
distributions of the cluster and field samples remain similar.
From the 1823 galaxies for which both \oii\ and H$\delta$ lie within the observed redshift range (excluding the
BCGs and cluster members beyond 2$R_{200}$),
we first select the 1572 which have $W_m\le 5$, so that the lowest quality data do not dominate our results;
this is equivalent to imposing a magnitude limit which varies from cluster to cluster.  
It is further desirable to remove those
galaxies from the sample which have extraordinarily large uncertainties in the measured
spectral indices, which often result from poor subtraction of bright
night sky lines.  The distribution of the error estimates for each index, scaled as discussed at the end of the
previous subsection, are shown in Figure \ref{fig-errordist}.  Based on these distributions, we have chosen to exclude
an additional 159 galaxies with errors greater than 15\AA, 5\AA, and 0.5 for \ow, \hd\ and D4000,
respectively.  This selection introduces another second order correction to $W_{\rm spec}$,
as galaxies with index errors exceeding our limit tend to be fainter.  We calculate, in bins of absolute
luminosity, the fraction of galaxies, weighted by $W_{\rm spec}$, removed from the
sample by this selection.  The remaining galaxies are weighted by the inverse of this number, which we
will call $W_{\rm err}$, multiplied by $W_{\rm spec}$.  $W_{\rm err}$ varies from 1 for galaxies with
$M_r<-19.8+5 \log{h}$, to $\sim2$ for galaxies with $M_r\approx -17.1+5\log{h}$.
The maximal sample consists of 1413 galaxies: 924 cluster, 407 field and 82 near field.
The absolute
magnitude distribution of the cluster sample (weighted by $W_{\rm spec}\times W_{\rm err}$) is compared with that of the field in
Figure \ref{fig-Mag_cf}.   Although the two distributions are similar, the cluster sample is
biased toward less luminous galaxies, and the weights do not sufficiently correct the spectroscopic
numbers (to be representative of the photometric sample) below $M_r=-18.8+5 \log{h}$, due to the removal of galaxies with $W_m>5$.

\subsection{The PEGASE Model Parameters}\label{sec-models}
As discussed in \S~\ref{sec-indices}, the \hd\ index is sensitive to its definition
and the manner in which it is measured.  Thus, its interpretation depends on
comparing it with models in which the index is measured in an identical
manner.  It is not possible to use model results published by
other authors (e.g., \cite{CS87,BP}) to interpret data unless it is shown that
the indices defined for those models are comparable to those defined for the data.  
In light of this, we will reconstruct models in \S\ref{sec-defs}
which have been explored by other authors, but based on our index definitions.

The evolution of \hd\ and D4000 with star formation history
for solar metallicity galaxies is obtained from the PEGASE population synthesis
code (\cite{PEGASE}), based on their UV to NIR spectral library.  
The PEGASE models
predict results similar to those of the updated GISSEL96 models (\cite{BC93}), but
provide more freedom in constructing spectra with complex star formation histories; in 
particular, different initial mass functions (IMFs) are readily implemented.  The drawback is that the models are
solar metallicity; to investigate metallicity effects, we will consider the GISSEL96 models.
The standard IMF adopted here is that of  Salpeter (1955\nocite{Sp}); 
the effect of adopting a truncated IMF model will be discussed in \S\ref{sec-variations}.

Spectral indices are determined from the model spectra in the same manner as from the
CNOC1 data; that is, the same code, with the same line definitions, are run on the spectra
output by PEGASE.   We have not included
Balmer--filling due to nebular emission; thus, the model \hd\ indices represent
an upper limit on the equivalent width one expects to observe when massive stars
are present. 

Care was taken to ensure that the line indices we obtain from the models correspond
to the line indices measured on the data; in particular, the \hd\ index is quite
sensitive to both the resolution and the sampling of the spectra.  The (rest frame) resolution of the
model spectra is 10\AA, comparable to the resolution of our
data (about 15\AA\ at z=0.3).  However, the model spectra are more coarsely sampled
than the CNOC1 data (every 10\AA\ instead of every 2.6\AA, rest frame), so 
we resample them to 2.6\AA,
interpolating using a fifth order polynomial fitting routine.  We verified that each
line index we measure on the CNOC1 spectra is unchanged if we first smooth the spectra
by averaging every four pixels, and then resample using our interpolation algorithm.

\section {Galaxy Classifications}\label{sec-defs}
There is, as yet, no clear consensus in the literature about what type of
galaxy is called, for example, a ``starburst'', ``H$\delta$--strong'', ``post--starburst'',
``E+A/K+A'', etc.  This is because the diagnostic indices are not
all defined in the same way, the spectra used vary in terms of S/N and resolution,
and the physical phenomena one is trying to isolate varies from author
to author.  In any case, such classifications are somewhat arbitrary, and there can be no
objectively correct choice.  
For reasons discussed in \S\ref{sec-compare}, we have chosen not to
use the most recent
classification scheme (Dressler et al. 1999), and thus wish to state clearly
how our classes are defined.  These should be compared and contrasted not
only with that paper, but also, for example, the differing definitions in Couch \& Sharples (1987),
Barbaro \& Poggianti (1996), Barger et al. (1996) and Morris et al. (1998).
In most cases, the differences in classification are ones of detail, and refined
precision of the evolving nomenclature.

\subsection{Determination of the H$\delta$ Threshold}\label{sec-hdthresh}
One of the main goals of this paper is to identify a population of galaxies
which were forming stars a short time ($< 1$ Gyr) ago, but have since stopped.
These must be distinguished from galaxies which are currently undergoing
star formation, and those which have not done so for a long time.  We will make
this distinction partly based on the spectra of local galaxies (see \S\ref{sec-defs-oiihd}),
and partly on model predictions, which is the purpose of this subsection. 

\begin{figure*}
\begin{center}
\leavevmode \epsfysize=8cm \epsfbox{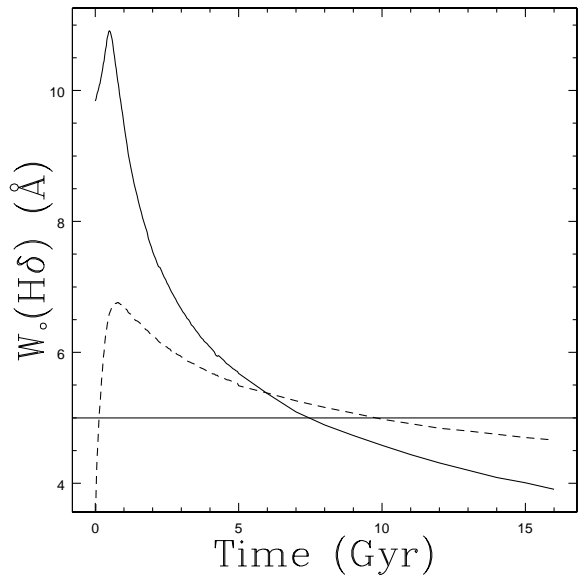}
\end{center}
\caption{The evolution of \hd\ with time, as determined from PEGASE models.
The {\it dashed line} traces the evolution of a galaxy with a constant star formation
rate, neglecting nebular emission, which will reduce \hd\ by $\sim1-3.5$ \AA.  
The {\it solid line} is the same model, but without stars more massive than 2.7$M_\odot$;
there will be no emission contribution in this case.
The horizontal line at \hd=5\AA\ is the lower \hd\ limit for our K+A galaxy
definition.  
\label{fig-Hdtime}}
\end{figure*}

In Figure \ref{fig-Hdtime} we show, as the dashed line, the evolution of the \hd\
index for a galaxy with a constant star formation rate.  As long as massive stars
are present, nebular emission filling of H$\delta$ will reduce the index by
between about 1\AA\ and 3.5\AA\ (\cite{BP}); thus, such a spectrum will not reach
\hd$\gtrsim$5\AA.  All normal, star--forming galaxies are therefore expected to have \hd$\lesssim5$\AA,
and nebular emission lines.  This is confirmed by the line indices of normal, spiral
galaxies, as we show in \S\ref{sec-defs-oiihd}.

To reach \hd$>5$\AA, the spectral light of a galaxy must be dominated by A--
and early F--type stars.  The presence of O-- and B--type stars, which have
weak intrinsic H$\delta$ absorption and can dominate the galactic light,
is the reason why the dashed line in  Figure \ref{fig-Hdtime} does not
reach \hd$>6.5$\AA, even without considering nebular emission contributions.
For illustrative purposes, we show the evolution of \hd\ in Figure \ref{fig-Hdtime}
for a model galaxy with a constant star formation rate, but with no stars more massive
than 2.7$M_\odot$ (corresponding
to an A0 star, B$\ddot{\mbox{o}}$hm--Vitense 1992).  There will be no emission 
contribution to such a spectrum, which therefore has \hd$>5$\AA\ 
for up to 7 Gyr; after that
the accumulation of less massive stars, which have weak \hd\ absorption, begins
to contribute significantly to the light. Clearly the absence of OB--stars allows
\hd\ to reach much higher levels.
As we will discuss in more detail in \S\ref{sec-defs-d4hd}, strong \hd\ indices
may be observed following the abrupt termination of star formation, due to 
the almost immediate death of the O-- and B--type stars.  This can be visualized
in Figure \ref{fig-Hdtime} as follows.  A constantly star--forming galaxy will
evolve at least $\sim1$ \AA\ below the dashed line; as soon as star--formation terminates, \hd\ will
increase approximately to the height of the solid line (which is an A--star dominated
spectrum).  For this reason, terminating star formation after 3 or 4 Gyr results in
only a weak increase in \hd, while terminating a short burst of a few hundred Myr
results in a much larger increase.   In either case, emission lines will
be absent in the spectrum.

Based on the results of these models, we adopt a threshold of \hd=5 \AA\ as
an acceptable division between galaxies with normal rates of star formation, and those
with recently truncated (or otherwise anomalous) star formation.  This threshold is
confirmed by the line indices of normal local galaxies, as discussed in the following subsection,
and may be lowered to \hd=3\AA\ for the reddest galaxies, as we demonstrate in \S\ref{sec-defs-d4hd}.

\subsection{Definitions Based on \ow\ and \hd}\label{sec-defs-oiihd}
\begin{figure*}
\begin{center}
\leavevmode \epsfysize=8cm \epsfbox{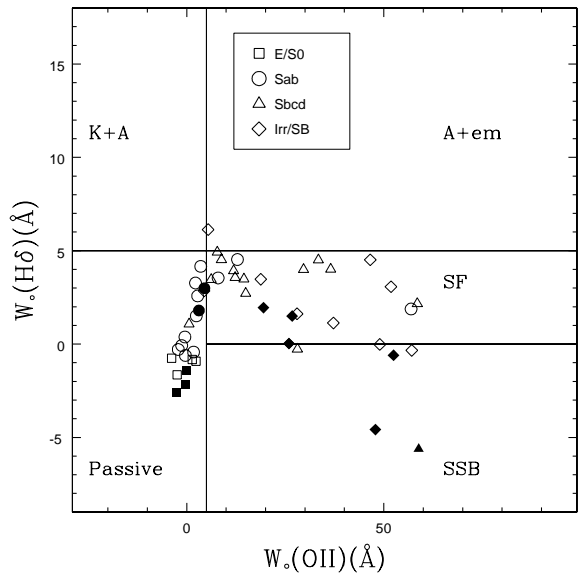}
\end{center}
\caption{Local data of Kennicutt (1992b, {\it open symbols}) and 
Kinney et al. (1996, {\it filled symbols}) presented in the \ow\--\hd\ plane. 
Positive values of \ow\ and negative values of \hd\ represent emission in these
indices. 
The plane is divided into five regions according to the definitions
in \S\ref{sec-defs-oiihd}:
Star--forming (SF) galaxies are those which have undergone significant star formation
for at least several hundred Myr, while star formation in short starburst (SSB) galaxies 
has begun within about 200 Myr.  The origin of A+em galaxies, which exhibit a strong
A--star spectrum with emission lines, is uncertain; some may be dusty starbursts, or host an AGN. 
 Of galaxies with \ow$<5$\AA, 
K+A galaxies may have recently had their star formation truncated, while those galaxies
with weaker \hd\ indices are evolving passively, with little or no star formation.  
\label{fig-HdOII_local}}
\end{figure*}

We will classify the galaxies in our sample based principally on their position 
in the \ow--\hd\ plane relative to the PEGASE models, 
and to local galaxies.  We show this plane in Figure \ref{fig-HdOII_local}. 
The data plotted are indices measured by running our
code with our line definitions on the local data of Kennicutt (1992b, open symbols)
and Kinney et al. (1996, filled symbols).  The different symbols represent the different
morphological classifications as shown on the figure.  The diamonds include any galaxy
classified as peculiar, irregular or starburst; all other points correspond to ``normal'',
local galaxies.  The plane is divided into five
regions, each of which we will  discuss separately.

\begin{itemize}
\item  \begin{bfseries} Passive Galaxy:\end{bfseries} This is a galaxy 
that is not currently undergoing significant star formation,
and thus evolves passively.  Spectrally, we identify such a galaxy as
one with \ow$<5$\AA, which is roughly our detection limit, and \hd$<5$\AA.
The local galaxies in this category are mostly elliptical, S0 or early type
spirals; the presence of these normal galaxies with \hd\ as large as nearly 5\AA\ justifies
the need for a threshold of at least this strength to clearly isolate galaxies with
anomalous star formation histories.

\item  \begin{bfseries}Star Forming (SF) Galaxy:\end{bfseries} This is a galaxy that has 
been undergoing significant star formation
for at least several hundred million years.  Unless heavily obscured by dust, the \oii\
emission line should be prominent, so \ow$>5$\AA.  Furthermore, models 
 show that $0<$\hd$<5$\AA; the lower limit is
determined from the results of Poggianti \& Barbaro (1996) and Barbaro \& Poggianti (1997), which
show that, if star formation has been ongoing for at least several hundred million years, the
stellar absorption will dominate any nebular emission of this feature.
Local galaxies that lie in this category are  both late Spirals (Sc) and
some galaxies that have been classified ``starburst'', based on the presence of strong 
emission lines.

\item  \begin{bfseries}Short Starburst (SSB):\end{bfseries} 
We define a short starburst as a {\em short lived ($<200$ Myr), large increase} in 
 star formation rate.  A galaxy that has had a
high star formation rate for more that a few hundred Myr will be considered a SF 
and {\em not} a SSB galaxy.  In this case, nebular emission will dominate the \hd\ feature and,
hence, both \ow\ and \hd\ will be emission features (\cite{PB96,BP}).

\item  \begin{bfseries}K+A Galaxy:\end{bfseries} This type of galaxy 
is one with no detectable \oii\ emission (defined as \ow$<5$\AA), but strong H$\delta$
absorption (\hd$>5$\AA).  Such a spectrum, which requires the presence of spectral type--A 
stars {\em without} more massive stars (which would give rise to [OII] emission lines),
may be the result of a recently terminated episode of star formation, most likely following
a short starburst (see, for example,
Dressler \& Gunn (1983), and  \S\ref{sec-defs-d4hd}).  Alternatively, some of these galaxies
may be extreme examples of the dust--obscured starburst scenario advocated by Poggianti et al. (1996)
to describe the A+em galaxies (see below).  This simple classification is based on only two
line indices, and our K+A galaxies do not comprise a completely homogeneous class in terms of
other spectral features, including the strength of other Balmer lines, or the \ion{Ca}{2} H and
K lines.  This must be kept in mind when comparisons are made with other work in which these features
are considered.

\item  \begin{bfseries}A+em Galaxy:\end{bfseries} Our final category consists
of galaxies with \ow$>5$\AA\ and \hd$>5$\AA.  This corresponds to a galaxy with
Balmer absorption lines as strong as those found only in A-- and early F--type stars,
but with \oii\ emission lines indicating the presence of more massive O-- and B--type
stars. There is not yet a clear interpretation of these spectra; the presence of OB--stars
should keep \hd$<$5\AA\ due both to emission--filling and the  weaker, intrinsic \hd\
absorption of these luminous stars.  It is possible that some of the observed emission
lines are due to an active nucleus; the useful diagnostic emission lines 
(H$\alpha$,H$\beta$,[\ion{N}{2}],[\ion{O}{3}]$\lambda$5007) are generally redshifted out of our
spectral range, so that we are unable to confirm this possibility.
Alternatively, Poggianti et al. (1999) suggest that these galaxies (which they call e(a)) may be dust--obscured starbursts;
we discuss this point further in \S\ref{sec-discuss}, and note that the same interpretation
may hold for the previously discussed K+A class of galaxy.
The lone point in the Kennicutt (1992b) sample which lies in this
region, shown on Figure \ref{fig-HdOII_local}, is NGC 3034, which is in fact a very dusty,
star--forming galaxy.  It has been classified ``starburst'' (e.g., \cite{Rieke}) based on
its strong H$\alpha$ emission, but
has long been known to have an integrated stellar spectrum analogous to A--stars (e.g., \cite{Humason});
this galaxy would thus seem to be of the type advocated by Poggianti et al. (1999).
\end{itemize}

\subsection{Additional Definitions Based on D4000}\label{sec-defs-d4hd}
To investigate the physical origin of the K+A galaxies, we will compare our
observations with the PEGASE spectrophotometric models 
to predict how \hd\ evolves with galaxy color, as determined
by the D4000 index.  This largely follows the method outlined by Couch \& Sharples (1987)
and Barger et al. (1996),
applied to the PEGASE models with our current index definitions.

\begin{figure*}
\begin{center}
\leavevmode \epsfysize=8cm \epsfbox{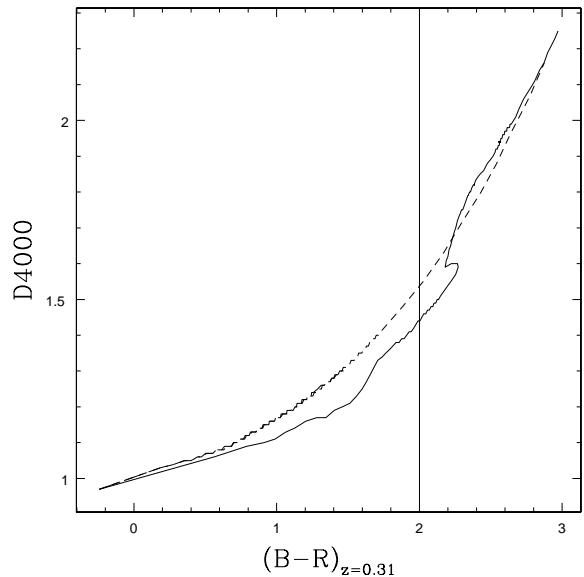}
\end{center}
\caption{Observed frame \br\ is plotted against rest frame D4000 for two different PEGASE
models.  The {\it solid line} represents the evolution
of a galaxy which formed all its stars in an initial, instantaneous burst.  The {\it dashed
line} corresponds to a galaxy with an exponentially decaying star formation rate, with
an e--folding timescale $\tau=2$ Gyr.  The vertical line represents the division adopted
by Barger et al. (1996), at \br=2; we find this corresponds to D4000$\approx 1.45$.
\label{fig-BRD4000}}
\end{figure*}

Barger et al. (1996) divided their sample into red and blue halves at \br=2 (observed at
$z=0.31$).  To convert this to an equivalent cut in D4000, we calculate
observed--frame ($z=0.3$) \br\ and rest frame D4000 for two different star formation histories, using the 
PEGASE models.  We show the results in Figure \ref{fig-BRD4000}; the models are described
in the caption.  From this figure, we adopt D4000=1.45 as the value corresponding to
\br=2; this value is fairly independent (within 0.05) of the model considered.

\begin{figure*}
\begin{center}
\leavevmode \epsfysize=12cm \epsfbox{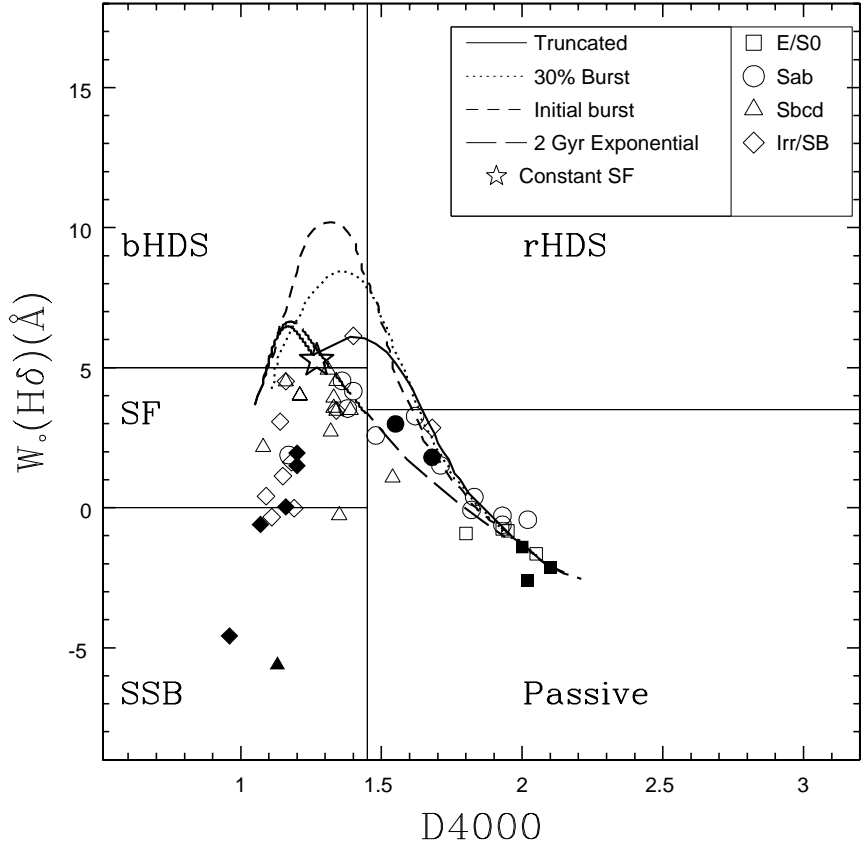}
\end{center}
\caption{PEGASE models in the D4000--\hd\ plane, for four different star formation
histories which are discussed in the Appendix.  Nebular emission contribution, not
included in the models, will lower the \hd\ index of any track with ongoing star formation, by
at least 1\AA.  
A galaxy with a constant star formation
is at the locus indicated by the {\it star}; an exponentially declining SFR gives rise to the
evolutionary track traced by the {\it long--dashed line}.  Truncating star formation in a galaxy with constant
star formation at 4 Gyr causes it to evolve along the {\it solid line}.  The {\it dashed line} shows an
initial, instantaneous burst of star formation, while the {\it dotted line} represents a 200 Myr
burst, involving 30\% of the stellar population of a constantly star forming galaxy (all
star formation is truncated following the burst).  
The plane
is divided into SSB (short starburst), rHDS (red H$\delta$--strong), blue HDS (blue H$\delta$--strong), SF (star--forming), 
 and Passive.  The local data are from Kennicutt (1992b, {\it open symbols}) and
Kinney et al. (1996, {\it filled symbols}). 
\label{fig-models_local}}
\end{figure*}

We now plot the output of five different models in the D4000--\hd\ plane, shown
in Figure \ref{fig-models_local} and described briefly in the caption; we defer detailed discussion of these models to the
Appendix.   The local data plotted on this
Figure are the same as those shown in Figure \ref{fig-HdOII_local}.
The plane is divided into five
regions similar to those defined by Barger et al. (1996), with the help of the D4000=1.45
division justified above.  We have also renamed some of the classifications
to correspond with our present terminology.  The Passive, SF and SSB regions are expected to
contain galaxies similar in star formation history to those of the same name defined in
\S\ref{sec-defs-oiihd}.  Galaxies with unusually strong \hd\ indices
are defined as either red or blue H$\delta$--strong (rHDS or bHDS); these will include both  K+A
galaxies and A+em galaxies.  Furthermore, normal galaxies in the red half of this plane generally
show little or no sign of star formation; both the models and real data of galaxies redward of D4000=1.45
have \hd$<3$\AA, and any contribution from nebular emission should be small.  Therefore, we lower
our \hd\ threshold to 3\AA\ for the rHDS galaxies, as only the truncated star formation models
pass through this part of the plane.

Both HDS categories will contain A+em galaxies, which have \ow$>5$\AA.
We discuss the suggestion by Poggianti et al. (1999), that these are dusty
starbursts, in \S\ref{sec-discuss}; 
whatever their origin, the A+em galaxies 
are not the  result of truncated star formation, since such activity is still present. 
To isolate those spectra which most likely result from truncated star formation, we define
two more galaxy classes, subsets of the rHDS and bHDS types:

\begin{itemize}
\item  {\bfseries Post-Starburst (PSB) Galaxies: }
We define these to be bHDS galaxies with \ow$<5$\AA. 
The PEGASE models shown in Figure \ref{fig-models_local} demonstrate that such a spectrum
requires the termination of a short burst of star formation.  Following this event, the 
galaxy spectrum will remain of the PSB type for about 300 Myr.  Truncated 
star formation without a burst (the solid line) causes the spectrum to
evolve quickly ($\lesssim100$ Myr) through this region, and during much of this time emission lines will be present, reducing
\hd\ to values less than 5 \AA.  Thus, few (if any) bHDS galaxies are expected
to result from truncated star formation without an initial starburst.

\item  {\bfseries Post-Star Formation (PSF) Galaxies:}
These are rHDS galaxies with \ow$<5$\AA, and their spectra can be matched with either
a recently terminated starburst episode, {\it or} through the abrupt cessation of
longer term star formation.  Galaxies evolving along either model track will spend about
300 Myr as a PSF galaxy; from the D4000 and \hd\ indices alone, it is not possible
to distinguish between these two possibilities.
\end{itemize}

\subsection{Summary of Definitions}
We have now defined six different types of ``unusual'' spectral types, based on models and the
indices of local galaxies.  All six of these types are rare, and can usually 
be matched by models in which star formation is truncated (T), sometimes following a short starburst (TSB);
the galaxies with emission lines are not matched by either of these models, and
are of uncertain
origin (U).  These six types, their definitions and interpretations are summarized in 
Table \ref{tab-typesum}.

\begin{deluxetable}{ccccc}
\tablewidth{0pt}
\footnotesize
\tablecaption{\label{tab-typesum}}
\tablehead{\colhead{Type}&\colhead{\ow}&\colhead{\hd}&\colhead{D4000}&\colhead{Interpretation}\nl
           \colhead{}&\colhead{(\AA)}&\colhead{(\AA)}&\colhead{}&\colhead{}}
\startdata
 K+A  &  $<5$ & $>5$ & --- & T,TSB\nl
 A+em  & $>5$ & $>5$ & --- & U\nl
 bHDS  & --- &  $>5$ & $<1.45$ & TSB,U\nl
 rHDS  & --- &  $>3$ & $>1.45$ & T,TSB,U\nl
 PSB   & $<5$ & $>5$ & $<1.45$ & TSB\nl
 PSF   & $<5$ & $>3$ & $>1.45$ & TSB,T\nl
\enddata
\tablecomments{Six spectral type classifications describing rare galaxy types which may
be the result of recently terminated star formation.  Column 5 lists compatible interpretations
of each spectral type: {\it T}: Truncated star formation without an initial starburst; {\it TSB}: Truncated
star formation following a burst; and {\it U}: Uncertain (due to the presence of emission lines).}
\end{deluxetable}

\section{Results}\label{sec-res}
\subsection{Spectral Index Dependence on Cluster--Centric Radius}
\begin{figure*}
\begin{center}
\leavevmode \epsfysize=15cm \epsfbox{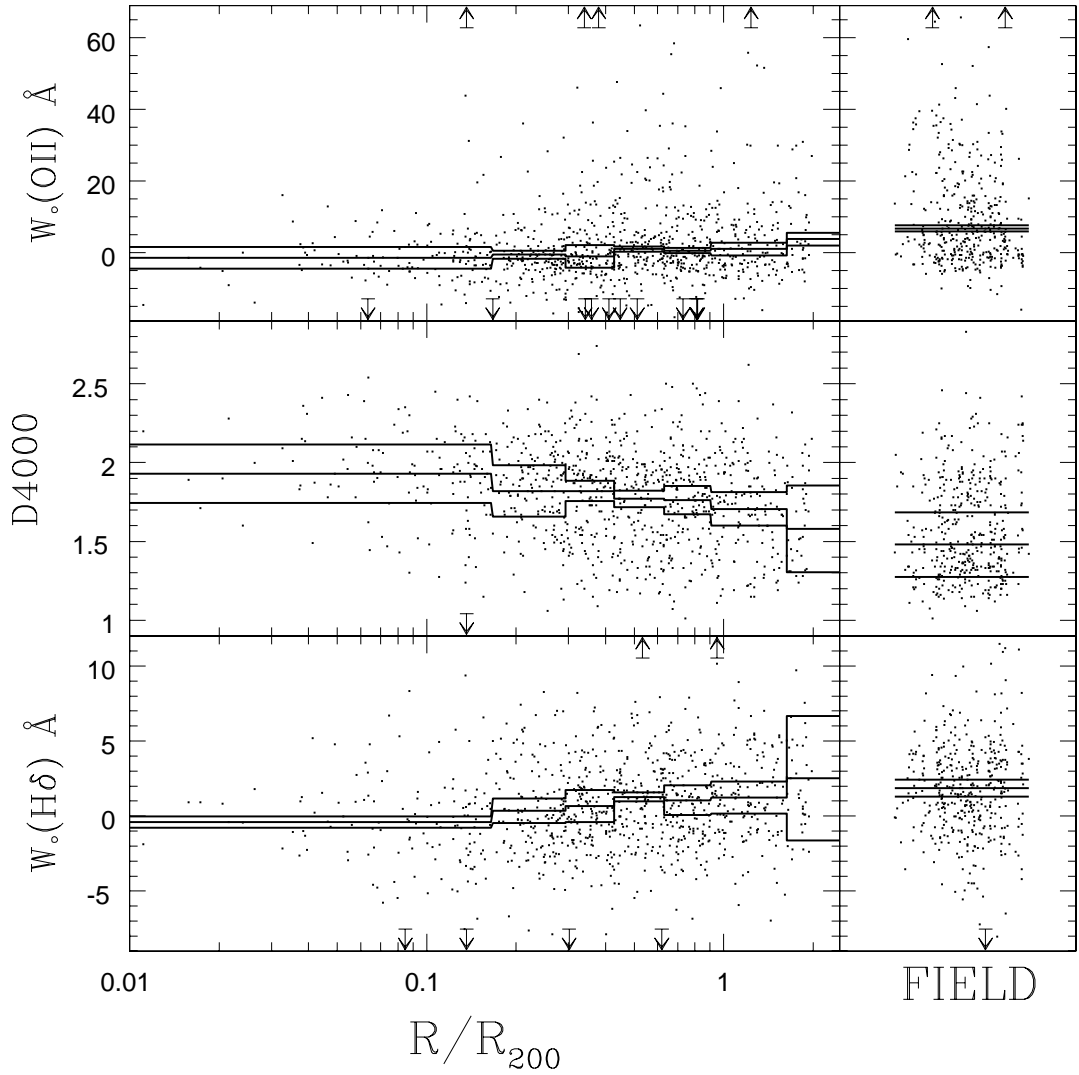}
\end{center}
\caption{Spectral indices plotted as a function of cluster--centric radius.  The sample
displayed in the right hand panels is the field galaxy sample, plotted against an arbitrary
abscissa for display purposes. 
Positive values of \ow\ and negative values of \hd\ represent emission in these
indices.  The {\it solid lines} are the median index
value for every 150 cluster galaxies, and the 
2$\sigma$ jackknife error estimates.  The range of data shown
is restricted to clarify the characteristics of the bulk of the data; arrows indicate
data which lie outside the plotted regions.
\label{fig-indrad}}
\end{figure*}

The measured values of \hd, D4000 and \ow\ in the maximal sample (\S\ref{sec-maxsamp}) 
are shown in Figure \ref{fig-indrad}, as a function
of cluster--centric radius, $R/R_{200}$.  Field galaxies are shown in the right hand
panel of each graph, plotted against an arbitrary abscissa.  
The solid lines in each graph are the weighted median value for
the field (right panel) and every 150 cluster points (left panel), with
2$\sigma$ jackknife errors (\cite{Jack1,Jack2}).  In the central regions, all three indices are inconsistent with the 
corresponding field value with at least 2$\sigma$ confidence, and they become
increasingly field-like with increasing radius.  The \hd\ index shows a 
small increase from the center of the cluster out to $R_{200}$, significant at about the 2$\sigma$
level; the decrease in D4000 over this range is measurable with about the same confidence.  The
median value of \ow\ is roughly constant throughout the cluster, within 2$\sigma$ limits, but clearly
the total distribution changes more strongly; particularly striking is the 
near total absence of galaxies with \ow$>20$\AA\ within
0.1 $R_{200}$, as shown previously in Balogh et al. (1997, 1998).
These trends are consistent with those observed in the individual clusters
Abell 2390 (\cite{A2390}) and MS1621.5+2640 (\cite{1621}), and are 
qualitatively consistent with a model in which most cluster galaxies evolve passively, and
the mean age of the galaxy stellar population decreases with increasing distance from the
cluster center.  

\subsection{Galaxies in the \ow--\hd Plane}\label{sec-OIIHd}
\begin{figure*}
\begin{center}
\leavevmode \epsfysize=12cm \epsfbox{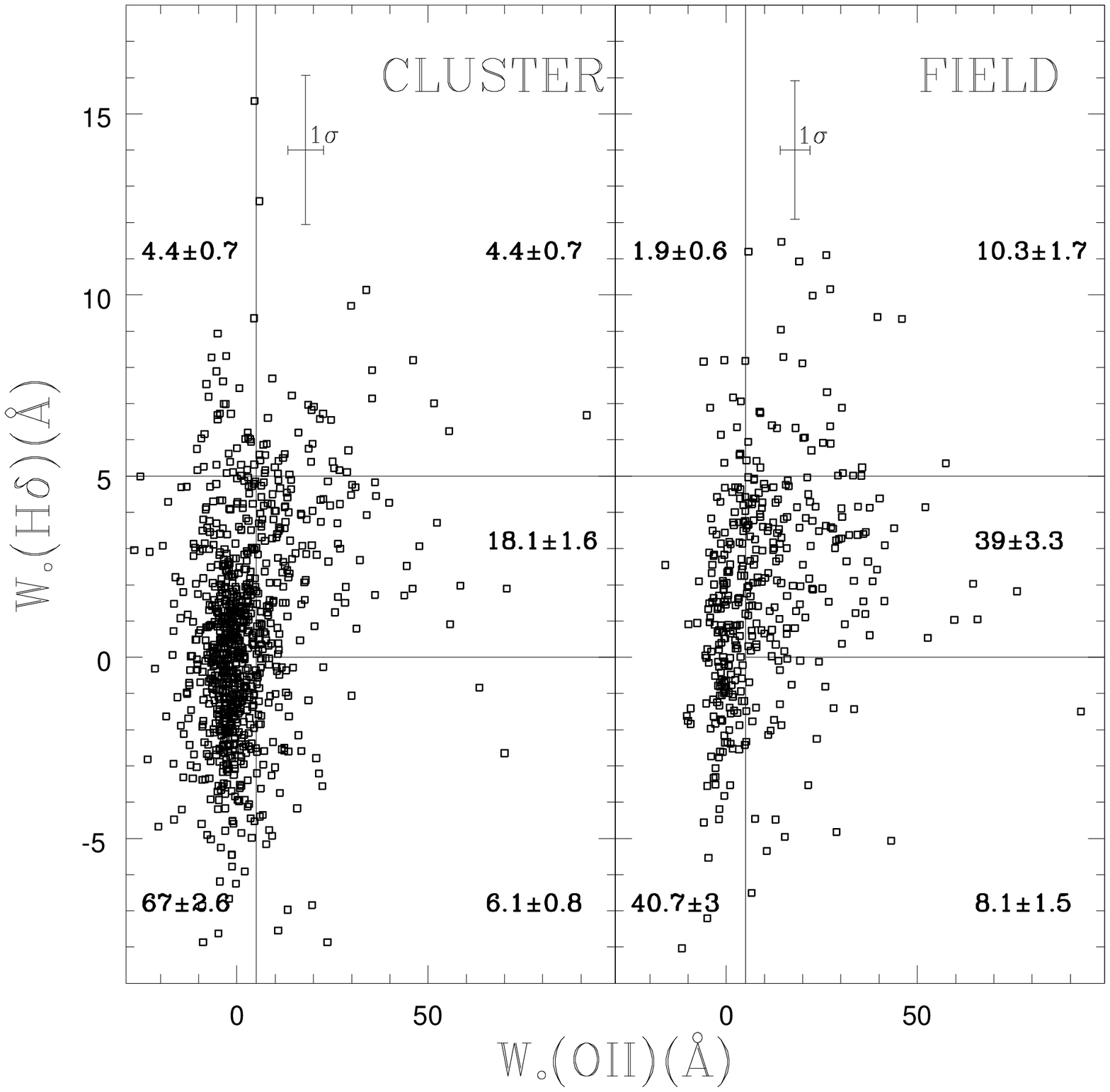}
\end{center}
\caption{The cluster and field data presented separately
in the \ow\--\hd\ plane, which is divided into regions defined
in Figure \ref{fig-HdOII_local} and discussed in \S\ref{sec-defs-oiihd}.  
Positive values of \ow\ and negative values of \hd\ represent emission in these
indices. 
The sample error bars represent the mean 1$\sigma$ error, and
the number in each region represents the weighted percentage of
galaxies in that region.  The errors assume
Poisson statistics, and do not account for the nonuniform scatter of data points
throughout the plane; see \S\ref{sec-scatter} for details of this effect.
\label{fig-HdOII}}
\end{figure*}

We plot \ow\ and \hd\ for the
cluster and field (maximal) samples separately, in  Figure \ref{fig-HdOII}.
Each panel is divided into the same
regions shown in Figure \ref{fig-HdOII_local} and discussed in \S\ref{sec-defs-oiihd}.   
The weighted percentage of galaxies
in each region is shown,  and the error associated with each number is
determined assuming Poisson statistics.  This error does not account for the uncertainty
due to unequal scatter in this figure; we will address this issue in \S\ref{sec-scatter}  

The fraction of galaxies undergoing {\em some} type of star--formation
(whether SF, SSB or A+em type) is $57\pm4\%$ in the field and only
$29\pm2\%$ in the clusters.  Despite the established B--O effect,
the mean star formation rate in these clusters is less than that in the 
field, as discussed in Balogh et al. (1997; 1998). 
There is an apparent excess of K+A galaxies in the clusters compared
with the field, significant at almost the 3$\sigma$ level.  As
we will show in the next subsection, most or all of this difference can be attributed 
to the scatter in this figure, and
to the slightly different absolute magnitude distributions of the cluster
and field samples.  In particular, the systematic effects of scatter result in an
overestimation of the K+A fraction; thus, we interpret the cluster fraction of
$4.4\pm0.7$\% as an upper limit.
This is consistent with the results
of other similarly defined samples at these redshifts, as we show in \S\ref{sec-compare}.  

\subsubsection{Determining the True K+A fraction}\label{sec-scatter}
\begin{figure*}
\begin{center}
\leavevmode \epsfysize=8cm \epsfbox{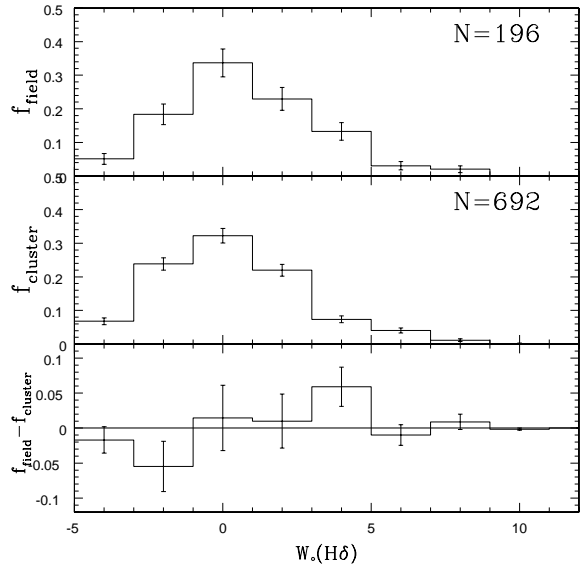}
\end{center}
\caption{The distributions of \hd\ for the subsample of galaxies with \ow$<5$\AA\
are shown for the field ({\it top panel}) and cluster ({\it middle panel}), normalized to unit area.  
The number of galaxies in each sample is shown in the top right corner; all error
bars are 1$\sigma$, assuming Poisson statistics.  The
difference between these two distributions is shown in the {\it bottom panel}.  There is no excess of cluster galaxies
with \hd$>5$\AA, which correspond to the K+A galaxies, relative to the field.
\label{fig-Hdhist}}
\end{figure*}

The large errors on both indices plotted in Figure \ref{fig-HdOII} spread the data out in
this plane; this results in non-uniform
scatter since galaxies from the most populated regions (Passive and SF) will preferentially
scatter into neighboring areas, biasing the fractions in these regions to larger values. 
In particular, the fraction
of K+A galaxies, which occupy the high-end tail of the \hd\ distribution, will be overestimated,
and we can expect this effect to be larger in the cluster sample, where the Passive galaxy
population is more dominant than in the field.  We can demonstrate this by assuming that
the true distribution of Passive galaxies is the same
in both the cluster and the field though the absolute number of such
galaxies in each environment is different.  Thus, the \hd\ distribution
of galaxies with \ow$<5$\AA\ should be the same if the fraction of K+A galaxies
is also identical.  In Figure \ref{fig-Hdhist}, we show these distributions normalized to unit area in the
top (field) and middle (cluster) panels.   Subtracting the cluster bins from the field bins yields the distribution
shown in the lower panel.  This allows easy comparison of the shapes of the \hd\ distribution
in the cluster and field samples: a relative excess of cluster galaxies with a given \hd\ would
result in a negative value in the bottom panel of this figure.  Instead, we see that 
the fraction of galaxies with \hd$>5$\AA,
corresponding to the K+A galaxies, is not significantly different in the two samples.  
That is, the difference between the (normalized) cluster and
field samples in the two bins with \hd$>$5 \AA\ is equal to zero within the 1$\sigma$
errors.  There is a 1$\sigma$ excess of field galaxies with \hd$\approx3$ \AA, which suggests
that even ``passive'' galaxies in the field may have somewhat younger stellar populations
than similar galaxies in clusters. 

We will now estimate a correction for the fractions in each of the regions of Figure \ref{fig-HdOII}.
The only way that a precise correction can be made is if the {\it a prior} distribution
is known, which is not the case here; therefore, some assumptions about this distribution
will have to be made.  We will consider two different methods of obtaining this correction:  the first is
applicable to each galaxy class, but requires stronger assumptions, while the second method
provides a more robust estimate but is only valid for the K+A galaxies.

\begin{figure*}
\begin{center}
\leavevmode \epsfysize=8cm \epsfbox{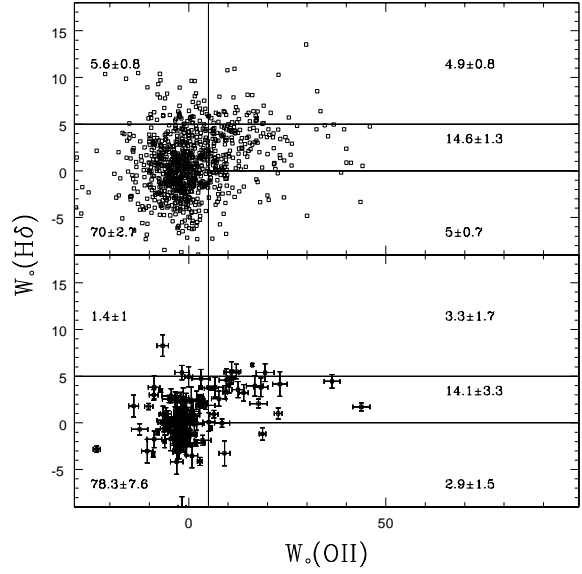}
\end{center}
\caption{The {\it bottom panel} shows cluster data with S/N$>5$, and errors in \hd\ and
\ow\ less than 1.5\AA\ and 2.5\AA, respectively.  Error bars are 1$\sigma$.  We add noise, characteristic
of the full sample error distribution, to each point, and generate twenty realizations
containing the same number of data points as the full cluster sample.  One such realization
is shown in the {\it top panel}.  The plane is divided into the same regions defined in
Figure \ref{fig-HdOII_local}; the numbers in each panel are the weighted percentage of objects
in that bin.  This experiment is used to calculate the scatter correction given in Table
\ref{tab-scatter}.
\label{fig-EAscatter_clus}}
\end{figure*}

\begin{figure*}
\begin{center}
\leavevmode \epsfysize=8cm \epsfbox{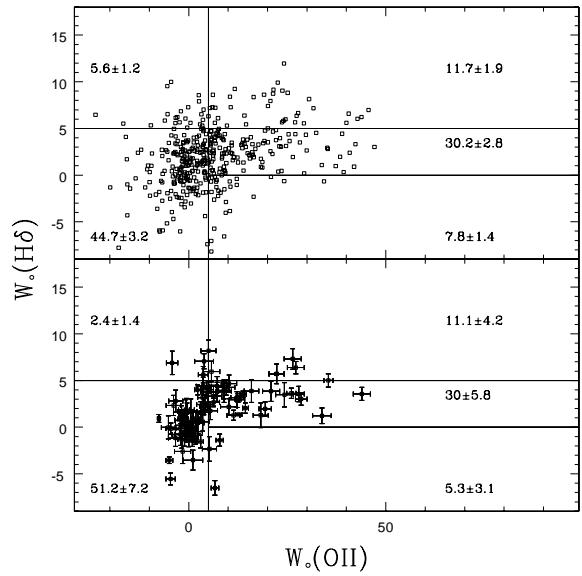}
\end{center}
\caption{This is the same as Figure \ref{fig-EAscatter_clus}, but for the field sample.
Again, the results are tabulated in Table \ref{tab-scatter}.
\label{fig-EAscatter_field}}
\end{figure*}

{\bfseries Method 1: Monte Carlo Simulations: }
We could estimate
the effects of scatter by adding appropriate noise to a ``pure'' sample with
no scatter.  We attempt to approximate such a sample by selecting galaxies
with reasonably high signal--to--noise
measurements (S/N$>5$) and errors in
both indices which are below the 25th percentile ($\Delta$\ow$<2.5$\AA\ and $\Delta$\hd$<1.5$\AA).
Admittedly, this is not quite what we want, as it is quite possible that this
subsample (generally with higher luminosities) is not distributed like the full sample;
however, it should give an approximately accurate correction, as confirmed by the second
method described below.
These data are shown in the bottom panel of Figures \ref{fig-EAscatter_clus} and
\ref{fig-EAscatter_field} for the cluster and field samples, respectively.

We now simulate noisy cluster and field samples by adding Gaussian random noise, with a variance
chosen randomly from the 1$\sigma$ uncertainties of the full (cluster or field, as
appropriate) data sample. 
We then generate several realizations, each one containing a number of points equal
to the number in the true cluster or field sample.  One such
simulation is shown in the top panels of Figures \ref{fig-EAscatter_clus} and
\ref{fig-EAscatter_field}.  From twenty realizations, we determine, for each galaxy class, the
average difference between the fraction of such galaxies in the 
noisy (simulated) and true (low--error sample)
distributions, and the variance of that difference.  We will refer to this value
as the ``scatter correction''.  


In Table \ref{tab-scatter}, we show the measured fraction of each galaxy type, from
Figure \ref{fig-HdOII}, in columns 2 (cluster) and 5 (field).  Columns 3 and 6 show the 
estimated scatter correction.  Subtracting these numbers from the
appropriate measured value, including the error in this correction, yields the corrected
fractions in columns 4 and 7.

\begin{deluxetable}{c|rrr|rrr}
\tablewidth{0pt}
\footnotesize
\tablecaption{\label{tab-scatter} Systematic Error Estimations in the \ow--\hd\ Plane}
\tablehead{\colhead{Object Type}&                  &\colhead{Cluster (\%)}&                  &                   &\colhead{Field (\%)}  & \nl
                     &\colhead{Measured}&\colhead{Scatter}  &\colhead{Corrected}&\colhead{Measured}&\colhead{Scatter}  &\colhead{Corrected}}
\startdata
 K+A     & $4.4\pm0.7$     & $2.8\pm0.5$  & $1.6\pm0.9$  & $1.9\pm0.6$  & $2.2 \pm0.9$  & $-0.3\pm1.1$\nl
 A+em    & $4.4\pm0.7$     & $1.4\pm0.5$  & $ 3.0\pm0.9$ & $10.3\pm1.7$  & $4.0\pm1.3$   & $6.3\pm2.1$\nl
 SF      & $18.1\pm1.6$    & $-2.4\pm0.6$ & $20.5\pm1.7$ & $39.0\pm3.3$ & $-3.3\pm 2.2$ & $42.3\pm4.0$\nl
 Passive & $67.0\pm2.6$    & $-7.1\pm1.0$ & $74.1\pm2.8$ & $40.7\pm3.0$ & $-6.2\pm1.4$  & $46.9\pm3.3$\nl
 SSB     & $6.1\pm0.8$     & $5.0\pm0.8$  & $1.1\pm1.1$  & $8.1\pm1.5$  & $3.4\pm 1.4$  & $4.7\pm2.1$\nl
\enddata
\end{deluxetable}

\begin{figure*}
\begin{center}
\leavevmode \epsfysize=8cm \epsfbox{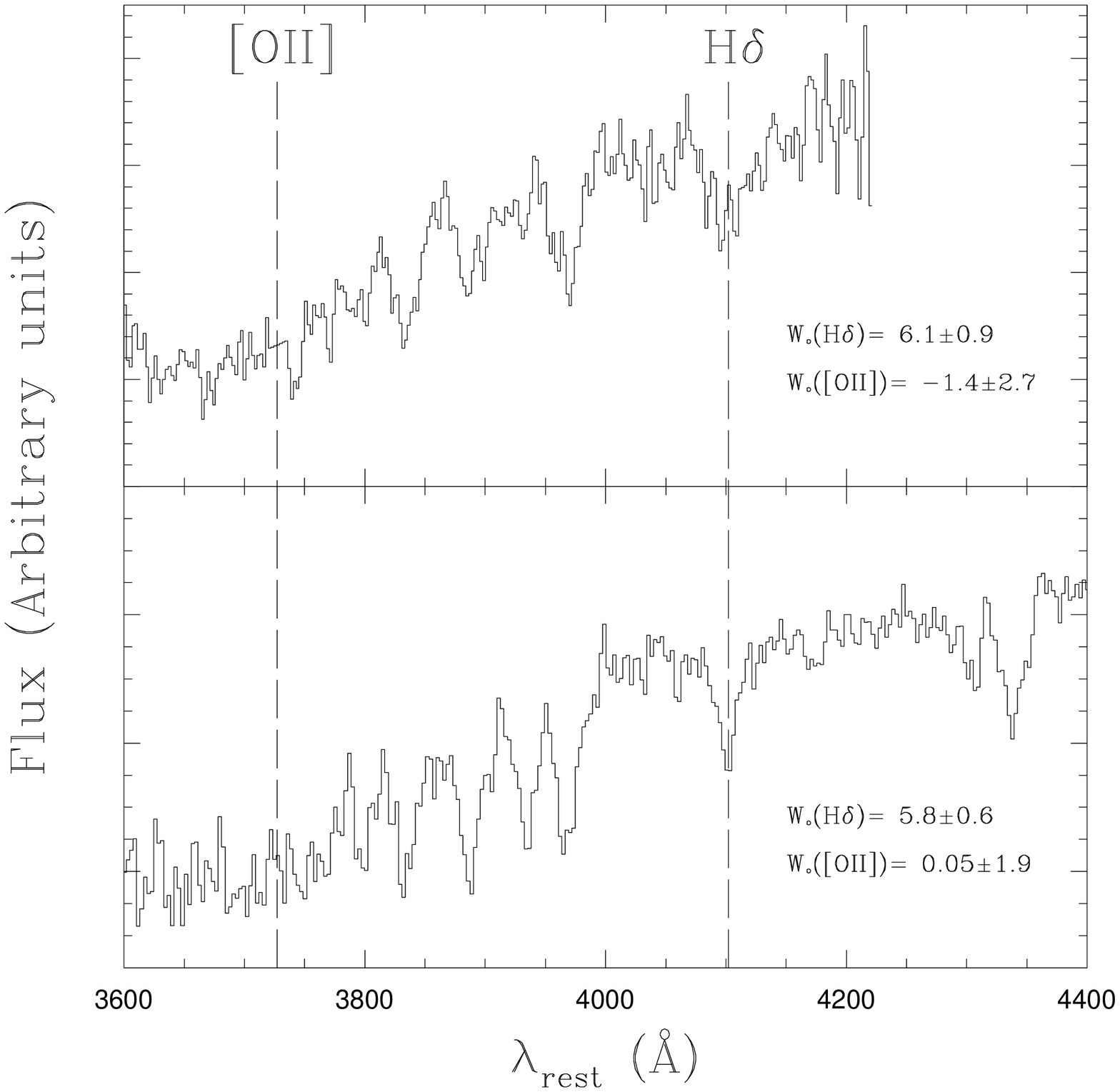}
\end{center}
\caption{Two high S/N examples of galaxies in our sample with a K+A type spectrum, in the
rest frame.  The {\it bottom panel} shows a cluster member in Abell 2390 at $z=0.2211$ 
(S/N=22, ppp \#100537);
the {\it top panel} shows a field galaxy in the field of MS1512 at $z=0.4926$ (S/N=7, ppp \#200607). 
These are good quality spectra, and not characteristic of typical spectra in the CNOC1 sample.
Note that, although both galaxies are classified as K+A, the top spectrum lacks the strong \ion{Ca}{2} K line
present in the bottom spectrum.
\label{fig-KAsample}}
\end{figure*}

We note that, in Table \ref{tab-scatter}, the corrected K+A fraction is equal to zero within 2$\sigma$
uncertainties.  However, 
this does {\it not} mean that K+A galaxies are not present in the sample; to provide concrete examples 
we show, in Figure \ref{fig-KAsample}, sample
spectra for two K+A galaxies: one found in the field and one in the cluster.  Note that the \ion{Ca}{2}
K line, prominent in the bottom spectrum, is weak/absent in the top spectrum; this is an example of two
galaxies which are assigned the same classification (K+A), though their spectra are not identical.

{\bfseries Method 2: Gaussian Cloud Analysis: }
An alternative way to determine the overestimate of the K+A fraction is to assume that it is
due only to scatter in the \hd\ direction from the Passive galaxies, i.e. those with \hd$<5$\AA\ and \ow$<5$\AA.
We can represent the \hd\ index of each galaxy as a Gaussian probability function,
with variance given by the 1$\sigma$ index uncertainty.  For each galaxy in the Passive sample,
we calculate the probability that \hd$>5$\AA, and then compute the weighted sum of all these probabilities.
  A smaller amount of ``backscatter'' (from the K+A
region to the Passive) will also
take place, and this must be subtracted off in an analogous manner.  This difference would give the total fraction of galaxies which have 
scattered from the Passive region into the K+A region, if the distribution of data in this plane was
representative of the error-free distribution; since the uncertainties significantly blur this 
distribution, this difference results instead in the {\it maximum} amount of K+A ``contamination'' expected.    
Following this procedure, we determine that
the maximum correction to the K+A fraction should be 2.8\% for the cluster, and 2.5\%
for the field (appropriate for the maximal data sample of \S\ref{sec-maxsamp}).  

We can now estimate the {\it minimum} correction required by counting the weighted number
of Passive galaxies with \hd$<-5$\AA.  Since the mean \hd\ in the Passive region is greater than
zero, and we expect few if any Passive galaxies to actually have \hd$<-5$, this fraction of
galaxies represents the minimum amount of scatter into the K+A region (without a backscatter
correction).  For the maximal cluster sample, this minimum correction is 1.9\%; for the field, it
is 1.2\%.

If we adopt the average of the minimum and maximum corrections, and half the difference to
represent (arbitrarily) a 2$\sigma$ uncertainty, we determine the best correction to
be 2.3$\pm0.25$\% for the cluster, and 1.8$\pm0.3$\% for the field, resulting in corrected K+A
fractions of 2.1$\pm0.7$\% and 0.1$\pm0.7$\%, respectively.  These corrections are somewhat
smaller than the ones in Table \ref{tab-scatter}, but equivalent within the uncertainties.  Although we take
this method to be the more robust estimate for the K+A correction, it confirms that the  
corrections listed in Table \ref{tab-scatter} are likely to be reasonable estimates.

\begin{figure*}
\begin{center}
\leavevmode \epsfysize=8cm \epsfbox{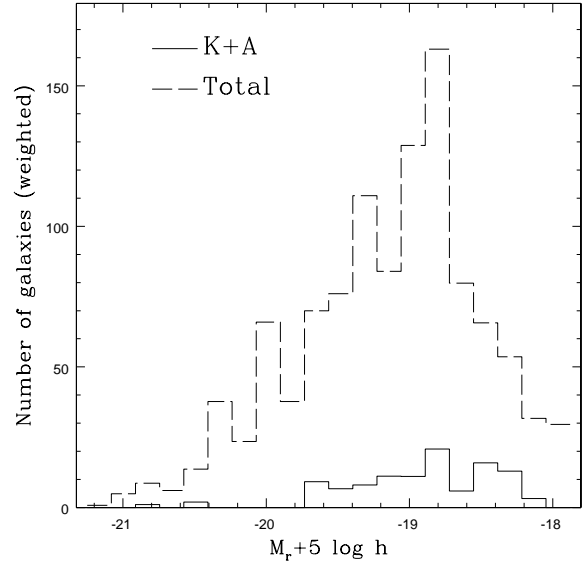}
\end{center}
\caption{The k-- and evolution--corrected absolute magnitude distribution for the K+A
galaxies ({\it solid line}) and the full maximal sample ({\it dashed line}).  Many of the K+A galaxies are 
less luminous than $M_r=-18.8+5 \log{h}$, where the statistical corrections are large and the maximal
sample is not representative of a complete photometric sample; the
abundance of K+A galaxies is therefore sensitive to the magnitude limit and statistical correction used.
\label{fig-KA_mag}}
\end{figure*}

\begin{figure*}
\begin{center}
\leavevmode \epsfysize=8cm \epsfbox{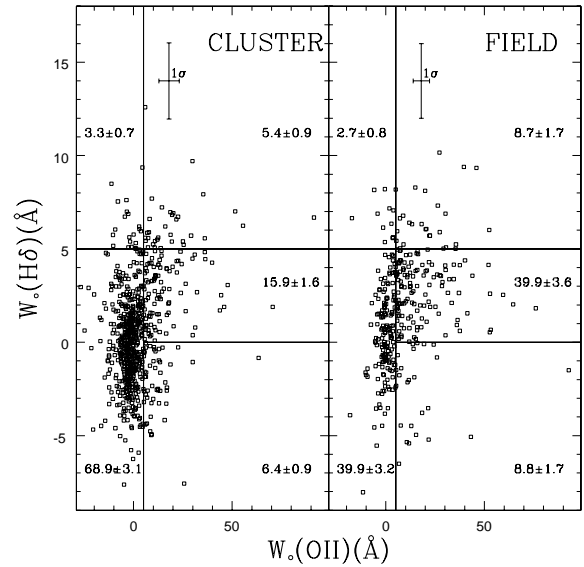}
\end{center}
\caption{The same as Figure \ref{fig-HdOII}, but for the luminosity limited
sample described in \S\ref{sec-maglim}.  The fraction of K+A galaxies in the cluster
and field samples are equivalent within 1$\sigma$ uncertainties.
\label{fig-Hdoii_mag}}
\end{figure*}

{\bfseries Luminosity Limit: }
In Figure \ref{fig-KA_mag} we show the $M_r$ distribution of the K+A sample, compared with the overall
distribution of the maximal sample.  This figure shows that many of the K+A galaxies lie 
below $M_r=-18.8+5 \log{h}$, where the magnitude weights are large, and the maximal sample is not
statistically representative of a complete photometric sample.  Thus, any absolute 
evaluation of the K+A fraction will be sensitive to the magnitude limit and statistical correction.  
In consideration of this,
we show the results for the luminosity limited sample (\S\ref{sec-maglim}) in
Figure \ref{fig-Hdoii_mag}.  For this restricted sample, the
fraction of K+A galaxies is equal to $3.3\pm0.7$\%
in the cluster, and $2.7\pm0.8$\% in the field; the two are equivalent within 1$\sigma$ uncertainties.
Adopting the Gaussian cloud scatter correction (Method 2), we find that these two numbers
are overestimated by 1.8$\pm0.4$\% and 1.5$\pm0.3$\%, respectively; thus, the true fraction is
reduced to 1.5$\pm0.8$\% in the cluster, and 1.2$\pm0.8$\% in the field.

{\bfseries Summary of Corrected Values: }
Although the amount of scatter correction necessary cannot be precisely determined (since the true
distribution of the data is unknown),
the uncorrected, cluster K+A fraction of 
4.4$\pm$0.7\% is a secure upper limit.  
Our best estimate of the true K+A fraction is thus obtained by applying the corrections determined
by the Gaussian cloud estimate.  For the maximal galaxy sample, we
find the fraction of K+A galaxies is 2.1$\pm0.7$\% in the cluster, and 0.1$\pm0.7$\% in the field.  
For the luminosity limited sample, the fraction is 1.5$\pm0.8$\% (cluster) and 1.2$\pm0.8$\% (field).
We conclude that, once scatter and luminosity effects are accounted for, there is no significant
difference between the frequency of K+A galaxies in the cluster and field environments.
We will compare these numbers with the results of other
surveys, both locally and at similar redshifts, in \S\ref{sec-compare}.

\subsubsection{The Radial Dependence}\label{sec-rad}
\begin{figure*}
\begin{center}
\leavevmode \epsfysize=8cm \epsfbox{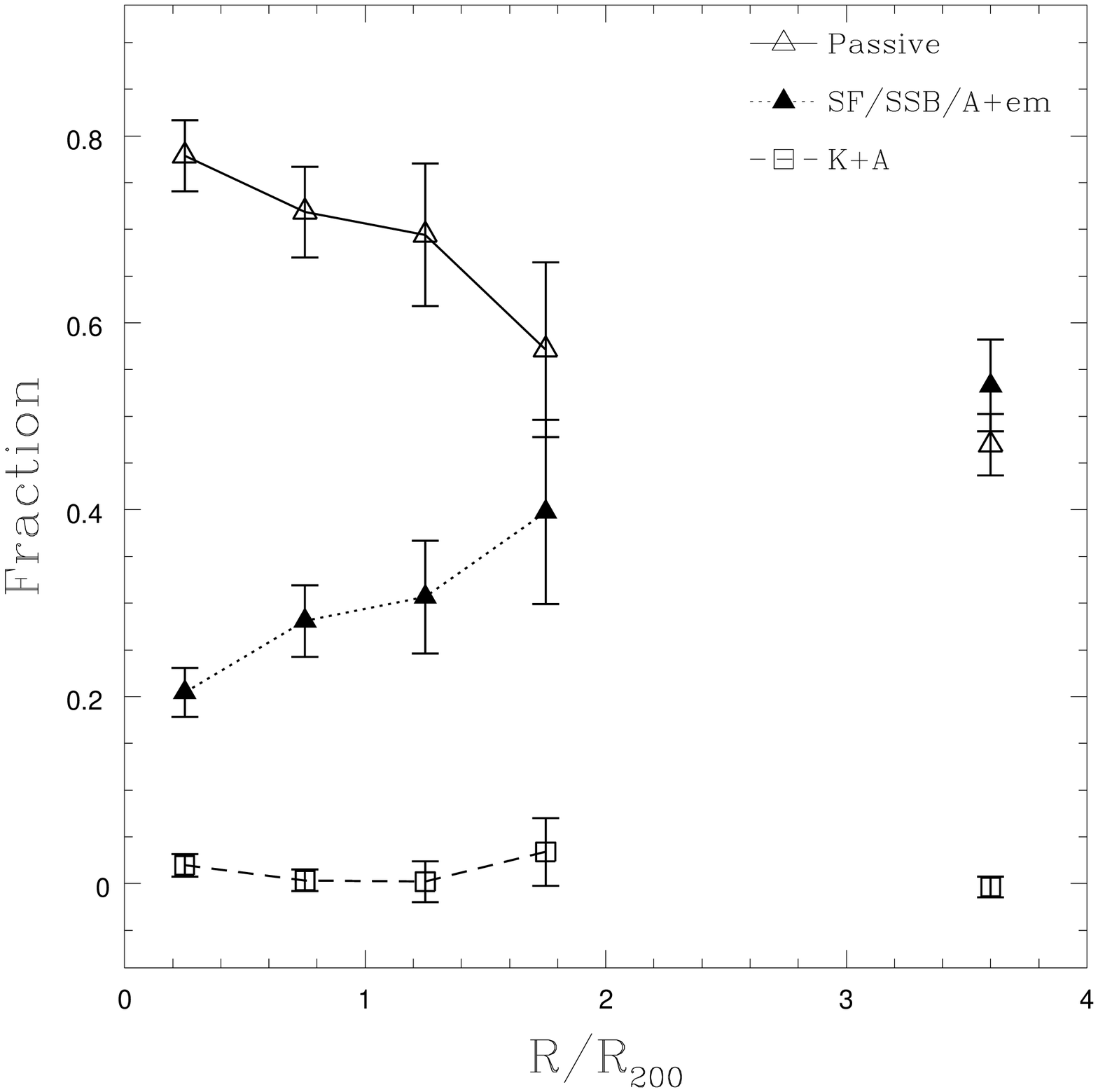}
\end{center}
\caption{The weighted fraction of each type of galaxy, as selected based on its position
in the \ow\--\hd\ plane shown in Figure \ref{fig-HdOII}, as a function of projected radius.
The fractions and 1$\sigma$ error bars are corrected for the systematic effects of nonuniform scatter, 
as discussed in \S\ref{sec-scatter}.
The connected points represent the cluster sample; the isolated points at
$R/R_{200}=3.6$ represent the values in the field sample, plotted at an
arbitrary radius for display purposes only.  
\label{fig-oiiHd_rad}}
\end{figure*}

The fraction of each galaxy type is plotted as a 
function of distance from the center of the cluster, normalized to $R_{200}$, in Figure \ref{fig-oiiHd_rad}.
The three ``active'' classes SF, A+em and SSB have been grouped together for clarity.
We have corrected each value for the estimated effects of scatter, using Method 1 
in \S\ref{sec-scatter}, and included the uncertainty of this correction in the error
bars.  
The galaxy population within the cluster exhibits a steadily increasing fraction of passively
evolving galaxies and a steadily decreasing fraction of star forming galaxies
with decreasing radius.  This is at least partly expected from the morphology--density
relation (Dressler 1980) but, as shown in Balogh et al. (1998) this relation cannot completely account
for the decrease in star formation rate toward the center of the cluster.  
The fraction of K+A galaxies within the cluster 
is never significantly in excess of the field value, at any radius.  

\subsubsection{The Redshift Dependence}\label{sec-z}
\begin{figure*}
\begin{center}
\leavevmode \epsfysize=8cm \epsfbox{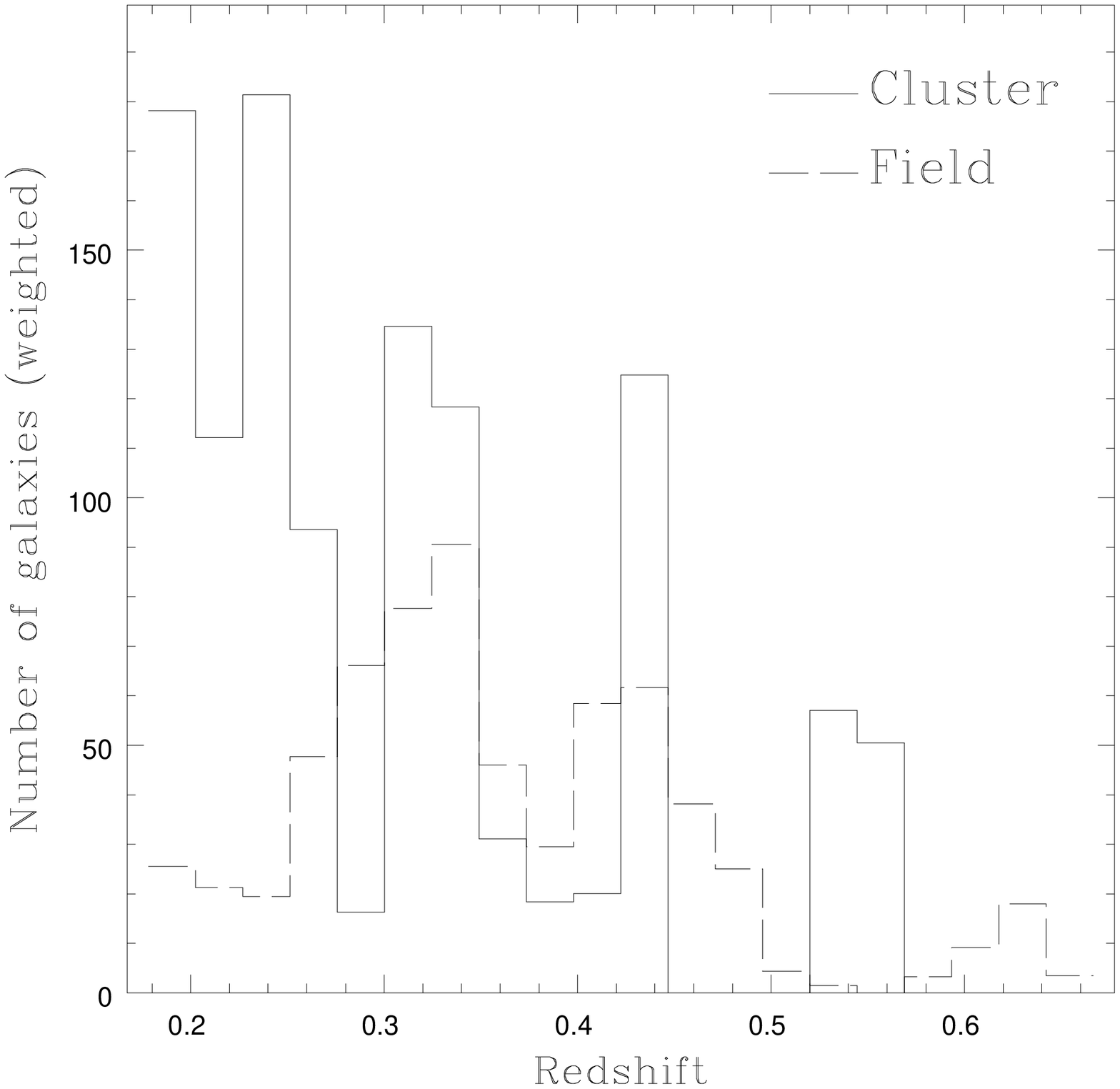}
\end{center}
\caption{
The redshift distribution of the cluster {\it (solid line)} and field
{\it (dashed line)} samples, limited to galaxies more luminous than
$M_r=-18.8+5\log h$ (corrected for band shifting and evolution).
 \label{fig-zdist}}
\end{figure*}

The fifteen clusters in the CNOC1 sample cover a large range in redshift, from
0.18 to 0.55 and, thus, it is of interest to investigate the possible redshift
dependence of the populations. To ensure an equitable comparison, we consider
the luminosity limited sample (\S\ref{sec-maglim}) here (recall that an evolution
correction is applied to the luminosities).
First, we show in Figure \ref{fig-zdist} the redshift distribution of this
sample, separately for the cluster and field.
This figure shows that the cluster observations are well divided into three,
well populated 
redshift bins: $0.15<z<0.28$; $0.28<z<0.35$; and $0.35<z<0.7$.  

\begin{figure*}
\begin{center}
\leavevmode \epsfysize=8cm \epsfbox{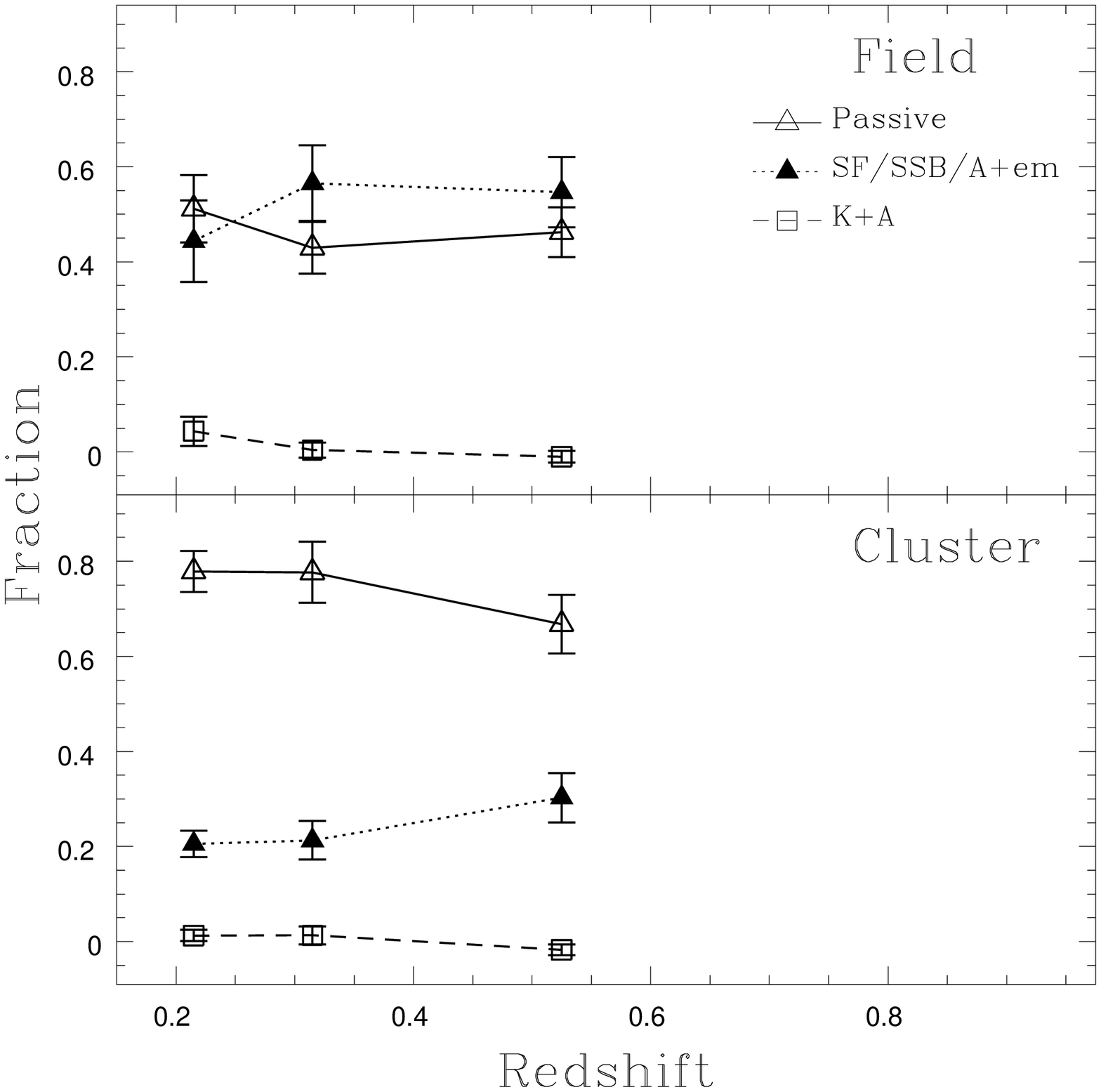}
\end{center}
\caption{This figure shows the fraction of each galaxy population defined in \S\ref{sec-defs-oiihd}
and Figure \ref{fig-HdOII}, as a function of redshift. Only galaxies more luminous than
$M_r=-18.8+5\log h$ are considered.  The field galaxy sample is shown
in the {\it top panel}, and the cluster sample in the {\it bottom panel}.  The 
statistically weighted data are corrected for
the systematic effects of scatter in the
\hd--\ow\ plane, as discussed in \S\ref{sec-scatter}.  All error bars are 1$\sigma$.
The K+A fraction is low in all three redshift bins.
\label{fig-HdOII_z}}
\end{figure*}

In Figure \ref{fig-HdOII_z} we plot the fraction of each galaxy type, as defined
in \S\ref{sec-defs-oiihd} and Figure \ref{fig-HdOII}, in each of the three redshift
bins described above.  Again,
we have grouped all star--forming galaxies together (SF, A+em and SSB) for clarity,
and corrected the values for the systematic effects of scatter using Method 1 of
\S\ref{sec-scatter}.
The cluster sample is shown in the bottom panel, and the field in the top panel.
A small increase in the fraction of star forming galaxies with redshift is observed,
though only significant at about the 1$\sigma$ level.
Importantly, the abundance of K+A galaxies does not show an increase
with redshift, which shows that the smallness of the fraction we observe is
not just due to the fact that our sample is dominated by low redshift $z\approx0.2$
galaxies.

\subsubsection{The Near--Field Galaxies}
\begin{figure*}
\begin{center}
\leavevmode \epsfysize=8cm \epsfbox{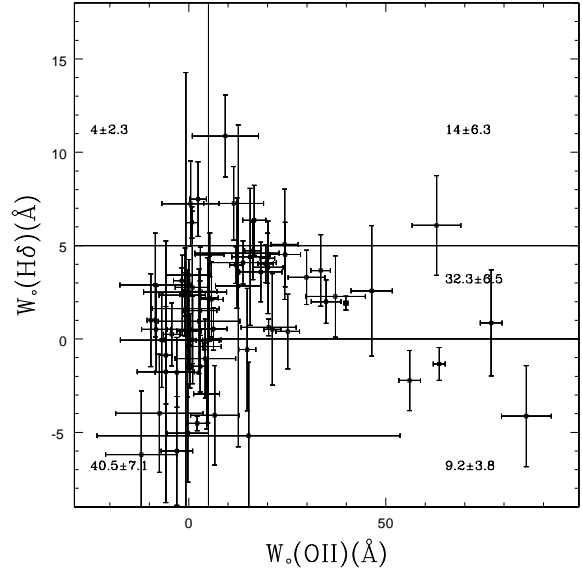}
\end{center}
\caption{Selected from the luminosity limited sample, we show 
the near--field galaxy population, which has velocities intermediate between our
cluster and field definitions, in the \ow\--\hd\ plane, with 1$\sigma$ error bars.  The regions are the
same as in Figure \ref{fig-HdOII_local}, and the number shown in each region represents the
weighted percentage of galaxies within that region.  The error is
estimated assuming Poisson statistics.  The fraction of galaxies in each class
is not significantly different from the field values shown in Figure \ref{fig-Hdoii_mag}.
\label{fig-HdOII_nf}}
\end{figure*}

In Figure \ref{fig-HdOII_nf} we show the \ow\--\hd\ relation
for the near--field galaxies in the luminosity limited sample (\S\ref{sec-maglim}).  The
near--field galaxies are a subset of galaxies with velocities intermediate 
between our cluster and field definitions (between 3 and 6 times the cluster velocity dispersion),
and will include both galaxies that are
far from the cluster ($\approx 50$ Mpc), with velocity differences that reflect the Hubble flow, and infalling galaxies
with peculiar velocities in excess of three times the cluster velocity dispersion. 
None of the galaxy abundances are significantly different from the field (c.f. Figure \ref{fig-Hdoii_mag}).  
Thus, even considering galaxies which may be in the infall regions of clusters, we do not
detect evidence for an excess of star forming or K+A galaxies, relative to the field.

\subsection{HDS and PSF Fractions}\label{sec-D4Hd}
\begin{figure*}
\begin{center}
\leavevmode \epsfysize=12cm \epsfbox{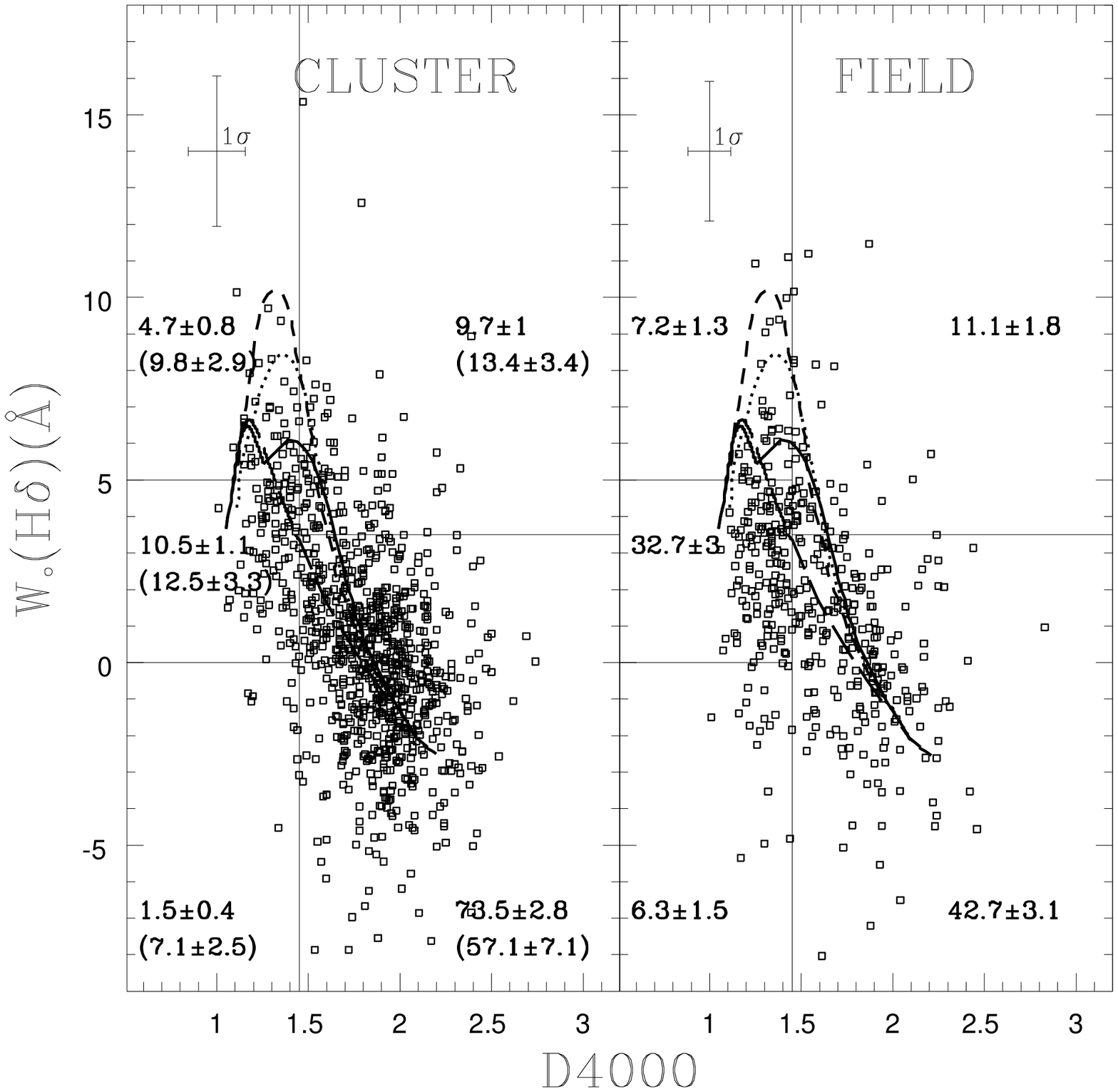}
\end{center}
\caption{The data presented in the D4000-\hd\ plane, 
divided into a field and cluster sample. The models and region delineations over-plotted 
are the same ones shown in Figure \ref{fig-models_local}.  The number in each region is the
weighted percentage of galaxies in that region, and the errors
assume Poisson statistics.  The numbers in brackets, for the cluster sample, are 
from Barger et al. (1996), corrected for the small differences in region definitions.
Our results are compared with those of Barger et al. in \S\ref{sec-comp-CS}.
The sample error bars represent the mean 1$\sigma$ error.
\label{fig-D4Hd}}
\end{figure*}

The measurements of \hd\ and D4000 for the maximal sample (\S\ref{sec-maxsamp})
are shown in Figure \ref{fig-D4Hd}, for the cluster and field, separately.
Each plane is divided into the five regions shown in Figure \ref{fig-models_local};
the number in each panel is the weighted percentage of galaxies within it.  
The numbers in brackets, for the cluster sample, are from Barger et al. (1996),
corrected for small differences in region definitions.  We will compare these numbers
in \S\ref{sec-comp-CS}.

We can correct the fraction of galaxies in each region of this figure for the effects of scatter in a 
manner analogous to the Monte Carlo method described in \S\ref{sec-scatter}; the results are shown in
Table \ref{tab-scatter2}.  After this correction, both 
bHDS (at 1.0$\pm$1.2\%) and rHDS galaxies (3.6$\pm$1.4\%) in clusters are not 
more common than they are in the field (2.7$\pm1.8$\% and 9.4$\pm$2.2\%, respectively).  

As discussed in \S\ref{sec-defs-d4hd}, the HDS galaxies include emission line galaxies;
populations with recently truncated star formation should fall into either the PSB or PSF subclasses.
In Table \ref{tab-HDS} we show, for the cluster and field separately,
the ``raw'' PSB and PSF measurement, the estimated scatter correction
(assuming that the  relative scattered fraction is the same  for the PSF/PSB classes
as for the rHDS/bHDS classes) and the corrected estimate of the ``true'' fractions.
Again, the fractions of PSB and PSF galaxies are not significantly more common in
clusters than they are in the field.  In both cases, the PSF galaxies are more
abundant, relative to the PSB galaxies, by a factor of $\sim 5-10$.  Though this
number is quite uncertain, if the large
excess is real it may have significant consequences, as we discuss 
in \S\ref{sec-discuss}.
\begin{deluxetable}{c|rrr|rrr}
\tablewidth{0pt}
\footnotesize
\tablecaption{\label{tab-scatter2} Systematic Error Estimations in the D4000--\hd\ Plane}
\tablehead{\colhead{Object Type}&                  &\colhead{Cluster (\%)}&                  &                   &\colhead{Field (\%)}  & \nl
                     &\colhead{Measured}&\colhead{Scatter}  &\colhead{Corrected}&\colhead{Measured}&\colhead{Scatter}  &\colhead{Corrected}}
\startdata
 bHDS    & $4.7\pm0.8$     & $3.7\pm0.9$  & $ 1.0\pm1.2$  & $7.2\pm1.3$  & $4.5 \pm1.3$   & $ 2.7\pm1.8$\nl
 rHDS    & $9.7 \pm1.0$     & $6.1\pm1.0$  & $ 3.6\pm1.4$  & $11.1\pm1.8$  & $1.7\pm1.3$    & $9.4\pm2.2$\nl
 SF      & $10.5\pm1.1$    & $-12.1\pm1.1$ & $22.6\pm1.5$  & $32.7\pm3.0$ & $-11.3 \pm 2.2$ & $44.0\pm3.7$\nl
 Passive & $73.5\pm2.8$    & $-0.8\pm1.2$ & $74.3\pm3.0$  & $44.7\pm3.1$ & $ 1.7\pm1.6$   & $43.0\pm 3.5$\nl
 SSB     & $1.5\pm0.4$     & $3.1\pm0.7$  & $-1.6\pm0.8$  & $ 6.3\pm1.5$ & $3.2\pm 1.4$   & $3.1\pm2.0$\nl
\enddata
\end{deluxetable}

\begin{deluxetable}{l|rrr|rrr}
\tablewidth{0pt}
\footnotesize
\tablecaption{\label{tab-HDS} Abundances of PSB and PSF Galaxies}
\tablehead{\colhead{Object Type}&                  &\colhead{Cluster (\%)}&                  &                   &\colhead{Field (\%)}  & \nl
                     &\colhead{Measured}&\colhead{Scatter}  &\colhead{Corrected}&\colhead{Measured}&\colhead{Scatter}  &\colhead{Corrected}}
\startdata
PSB & 1.3 $\pm$0.4  & 1.0$\pm$0.2 & 0.3  $\pm$0.4& 0.7 $\pm$0.3  & 0.4$\pm$0.1 & 0.3 $\pm$0.3 \nl
PSF & 4.9 $\pm$0.6  & 3.0$\pm$0.5 & 1.9 $\pm$0.8 & 3.6$\pm$0.9  & 0.5$\pm$0.5 & 3.1 $\pm$1.0 \nl
\enddata
\end{deluxetable}

\subsubsection{Reddening, Metallicity and IMF Effects}\label{sec-variations}
A striking feature of Figure \ref{fig-D4Hd} is the abundance of 
galaxies, both cluster and field, with D4000 $\gtrsim1.9$ and \hd\ $\gtrsim2$\AA, which are not matched by
any of the models in Figure \ref{fig-models_local}.  This problem has also been pointed out by
Couch \& Sharples (1987), Morris et al. (1998) and Poggianti \& Barbaro (1996), among others, 
and persists if photometric colors
are considered instead of D4000.  The problem has been substantially alleviated by considering
the narrower D4000 index, but it still persists.  

\begin{figure*}
\begin{center}
\leavevmode \epsfysize=12cm \epsfbox{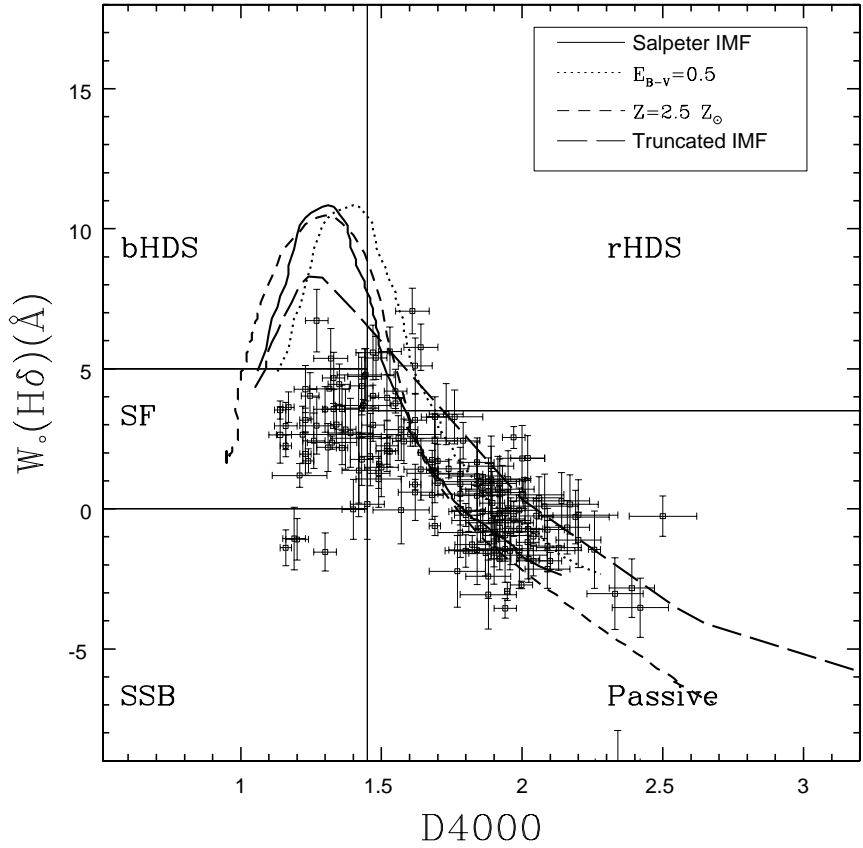}
\end{center}
\caption{PEGASE star formation models presented in the D4000--\hd\ plane, with
regions delineated as in Figure \ref{fig-models_local}.  The {\it solid line} is the standard, Salpeter
IMF initial burst model, also shown in Figure \ref{fig-models_local}.  This model, reddened
by \ebv=0.5, is shown as the {\it dotted line}.  A truncated IMF initial burst (Rieke et al. 1993) is shown as
the {\it long--dashed line}, and a high metallicity (GISSEL96) model is the {\it short--dashed line}.  
The data plotted are only those with high 
signal--to--noise ratios (S/N$>$15) and uncontaminated by bright night sky lines; 
the error bars are 1$\sigma$. 
\label{fig-variations}}
\end{figure*}

In Figure \ref{fig-variations}, we plot only the data  with S/N $>$15 and uncontaminated
by bright night sky lines near the \hd\ and \ow\ indices.
For reference, we plot the initial burst, Salpeter IMF model shown in Figure \ref{fig-models_local},
with a solid line. Using the {\em deredden} task within IRAF, we redden this model 
by a large amount,  \ebv=0.5; this moves the model to the right by
about 0.15, as represented by the dotted line.  This reddened model extends 
in D4000 nearly to the full extent of the data, and the scatter in \hd\ at the red end is fairly uniform about
it.

We also consider 
two other model variations.  The short--dashed line in Figure \ref{fig-variations} represents an initial burst model for a 
galaxy with super-solar metallicity, Z=2.5$Z_\odot$,  generated from the updated Bruzual
\& Charlot (1993) models.  Although this model does indeed extend to very large D4000, the 
\hd\ index is always negative\footnote{Recall, from \S\ref{sec-indices}, that a negative \hd\ index
does not necessarily imply an emission feature.} for D4000$>1.9$ and, thus, lies below the bulk of the data.
Secondly, we consider the effects of an extreme IMF model, after Rieke et al. (1993), in which
no stars less massive than 2$M_\odot$ are formed.  An initial burst of star formation with
this IMF is shown in Figure \ref{fig-variations} as the long--dashed line.  
This model is also able to match the strongest \hd\ indices of our data
with D4000$>1.9$.  However, the
stellar population resulting from such a burst is very short lived, about 300 Myr;
thus, it is unlikely that many of these red galaxies can be
undergoing such a burst, and we expect that significant dust reddening
is the best explanation of the data.  This supports the finding of Poggianti et al. (1999),
that dust obscuration in galaxies at these redshifts plays an important role in the
appearance of their spectra.

\section{Comparison With Previous Work}\label{sec-compare}

\subsection{The Low Redshift Universe}
The Las Campanas Redshift Survey (LCRS),  as presented
in Zabludoff et al. (1996), is a large ($\sim 24,000$ galaxies), unbiased sample of relatively nearby
galaxies ($0.05<z<0.13$)
for which good quality spectra and line indices are available.  Zabludoff et al.  identify 21 of their
11,113 galaxy subsample ($M_r\gtrsim -19+5\log{h}$) as E+A (renamed here K+A) types, based on \ow\ and an average equivalent width of
H$\delta$, H$\gamma$, and H$\beta$ ($<H\delta\gamma\beta>$).  They estimate that 1785 galaxies
in their sample lie within the infall radius of a rich cluster, and five of these are identified
as K+A types; therefore, $0.28\pm0.12$\% of the cluster population
are K+A galaxies, compared with $0.17\pm0.04$\% in the field.  These fractions are equivalent
within 1$\sigma$ uncertainties.  The definition of a K+A galaxy adopted by Zabludoff et al.
differs slightly from the one we have adopted here (\S\ref{sec-defs-oiihd}) in two respects.  First, they use a minimum
Balmer index of $<H\delta\gamma\beta>$=5.5\AA.  From the PEGASE models we find that, for \hd$\approx 5$\AA, 
 $<H\delta\gamma\beta>-$\hd$\approx 0.7$; thus, our limit of \hd$>$5 \AA\ is nearly equivalent to theirs.
Secondly, Zabludoff et al. require \ow$<2.5$\AA; our lower S/N requires us to use a more
generous definition of \ow$<5$\AA.  Adjusting for this difference allows the inclusion of an additional 
$\sim 13$ K+A galaxies in the LCRS sample, raising the
total K+A fraction to $0.30\pm0.05$\%.  Since the uncertainties on the LCRS line indices are small
(due to the high S/N of the data), a scatter correction like the one discussed in \S\ref{sec-scatter}
is not made.

In our luminosity limited sample (\S\ref{sec-maglim}), which most closely matches the
luminosity range of the Zabludoff et al. sample,
the ``raw'' K+A fraction in the field is 2.7$\pm0.8$\%, which we regard as an upper limit.  Our best
estimate of this fraction, corrected for scatter using the Gaussian cloud method (\S\ref{sec-scatter}),
is 1.2$\pm$0.8\%.  Thus, the evolution in the K+A fraction over this fairly small redshift
range (from $z\sim0.1$ to $z\sim0.3$) amounts to a factor of only about 4$\pm$2.7, and the two fractions
are actually consistent at about the 1$\sigma$ level.

\subsection{Couch \& Sharples (1987) and Related Work}\label{sec-comp-CS}
Much of the original work in this field was done by Couch \& Sharples (1987)\nocite{CS87}, whose dataset
of 152 galaxies more luminous than about $M_r=-20+5\log{h}$ in three clusters at $z\approx0.31$ was later
analyzed in more detail by Barger et al. (1996).  As shown in Figure \ref{fig-D4Hd},
we find somewhat fewer rHDS and bHDS galaxies than Barger et al., though the difference is significant
at less than the 2$\sigma$ level, and may be partly due to an imprecise mapping of \br\ to D4000 (which
was determined in \S\ref{sec-defs-d4hd}).  
Barger et al. performed a useful and detailed analysis of the distribution
of galaxies in this figure, and compared it with model simulations to determine the
importance of starbursts to cluster galaxy evolution.  They suggest the data are well represented
by a model in which $\sim$ 30\% of cluster galaxies have undergone a $0.1$ Gyr starburst
(which produces 10--20\% of the the galaxy's final stellar mass) in the last $\sim$2 Gyr.
Given the small number of galaxies and the considerable uncertainties on the line indices,
these numbers are uncertain, as the authors acknowledge.  In \S\ref{sec-discuss}
we will argue from the fraction of K+A galaxies in the CNOC1 sample that the importance
of such short--lived starbursts in clusters is unlikely to be this large.

The Couch \& Sharples data are also considered by Poggianti \& Barbaro (1996), who conclude
from their modeling that strong starbursts are required to match the K+A galaxies in
that sample.  However, the comparison between the data and models may be
compromised by a difference in index definition.  The Couch \& Sharples data have
\hd\ indices as large as 9 \AA, with many points around $5<$\hd$<6$ \AA, while the
post--starburst models of Poggianti \& Barbaro struggle to reach \hd$=6$ \AA\ (see their fig. 4), and then only for short
periods of time.  By dividing the data and models into broad categories (i.e. \hd$>3$ and $(B-R)<2$)
it is possible to misinterpret the results if the models do not match the data {\it within}
a given category, as may be the case here.

\subsection{Cl1358+62 and CFRS}
Fisher et al. (1998) analyzed spectra for 232 galaxies within a $\sim$4 Mpc region around
the $z\approx0.32$ cluster
Cl 1358+6245, with a magnitude limit of $M_r\approx-19.1+5\log{h}$.  This improves on the preliminary work presented by Fabricant
et al. (1991), which used only 70 spectra in the central regions of this cluster.
These authors define K+A (which they term E+A) galaxies in a manner similar to 
Zabludoff et al. (1996), but with more generous limits in $<H\delta\gamma\beta>$ and
\ow.  
They find 11 (4.7$\pm1.9$\%) of their cluster galaxies show K+A
type spectra, remarkably similar to our total cluster fraction, before the necessary scatter 
correction.  Since
their uncertainties are also comparable to ours, we expect this correction also to be similar, 
resulting in a reduced, true K+A fraction.  

A similar comparison can be made
with the work of Hammer et al. (1997), who find a K+A fraction of 4.9$\pm1.5$\% in the
CFRS field galaxy sample,  with a less precise K+A definition, roughly similar to that of Fisher et al.  
Again, this number should be reduced due to the overestimation resulting from unequal scatter
and, thus, 
the amount of evolution implied relative to the low redshift (LCRS) sample will be significantly
less than the order of magnitude implied by the raw fraction alone.  This number is similar
to our raw cluster K+A fraction, but somewhat larger than our field value.
Thus, it supports our conclusion that K+A galaxies are not preferentially found in clusters;
however, we note that the mean
redshift of the CFRS, $z\approx0.6$, is considerably higher than that of the CNOC1 sample, and
evolutionary effects may be important. 

\subsection{The MORPHS Collaboration}\label{sec-morphs}
After this paper was submitted,
the MORPHS collaboration published a similar study based on ten clusters at $0.37<z<0.56$, including
657 galaxies brighter than $M_r\approx-19+5 \log{h}$.
(Dressler et al. 1999\nocite{D+99}; Poggianti et al. 1999\nocite{P+99}).  These authors conclude
that galaxies without detectable [\ion{O}{2}] emission but \hd$>3$\AA\ (which they call k+a/a+k)
are significantly more common in the cluster sample than the field, (21$\pm2$\% compared with
6$\pm$3\%).  Both the largeness of these fractions, and the significance of their difference,
are apparently in conflict with the results we have presented in \S\ref{sec-OIIHd}.  This difference
warrants a detailed comparison of the data and analysis, which we present in this subsection.

We will first show that our data are of comparable quality to the MORPHS data,
and then suggest that the differences in our conclusions may result from (1) different selection criteria
(both of galaxies and of clusters);
(2) different treatment of uncertainties; and (3) differences in galaxy classifications
and definitions.

\subsubsection{A Comparison of Data Quality}\label{sec-morphs-data}
\begin{figure*}
\begin{center}
\leavevmode \epsfysize=8cm \epsfbox{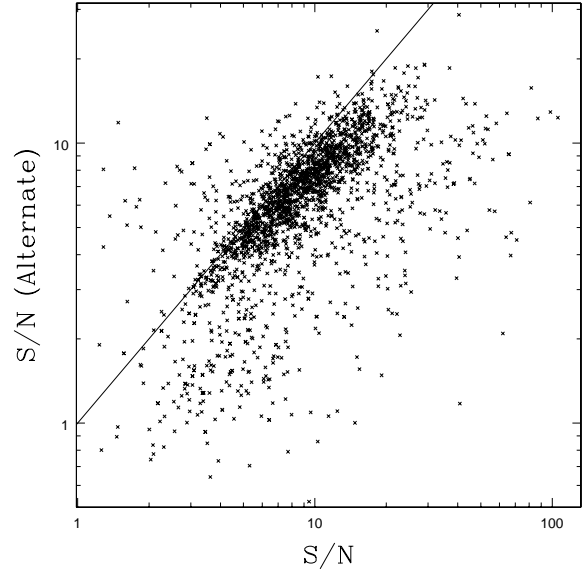}
\end{center}
\caption{Using the CNOC1 sample, we compare the S/N measured in the manner defined
in \S~\ref{sec-cnocsample} (on the x-axis) to the alternate definition described in \S~\ref{sec-morphs-data}.
The {\it solid line}, drawn for reference, is where the points would like if both
S/N measurements were equal.  Most of the data show a
strong correlation, though there is significant scatter.
The alternate definition underestimates the S/N, especially at high values, where real features in the continuum
add to the {\it r.m.s.} which is used instead of the noise.  
\label{fig-SNvsSNalt}}
\end{figure*}

We have obtained the publicly available MORPHS data from
http://www.ociw/$\sim$irs, and computed our indices from their data directly. 
However, the MORPHS data do not include error vectors, and uncertainty 
estimates of their indices are unpublished; thus care must be taken before quantitative comparisons
are made with our results.
In particular, continuum S/N ratios cannot be computed in the same way as we have done in \S\ref{sec-cnocsample}.  Instead,
we define an alternate estimate as the mean flux per pixel divided by the {\it r.m.s.} in the
wavelength region $4050<\lambda/$\AA$<4250$.  This underestimates the true S/N, especially for high S/N spectra, 
due to real features in the spectrum.  In Figure \ref{fig-SNvsSNalt} we show
how this ``alternate'' value of S/N, computed for the CNOC1 galaxies, compares with our definition in 
\S\ref{sec-cnocsample}.  The correlation is evident, though the scatter is large, and the
underestimation of S/N at high S/N values is clear.  Based on this correlation, we will assume
 that our alternative definition can hold as a reasonable proxy for the more realistic 
measurement used to qualify the CNOC1 data.  

\begin{figure*}
\begin{center}
\leavevmode \epsfysize=8cm \epsfbox{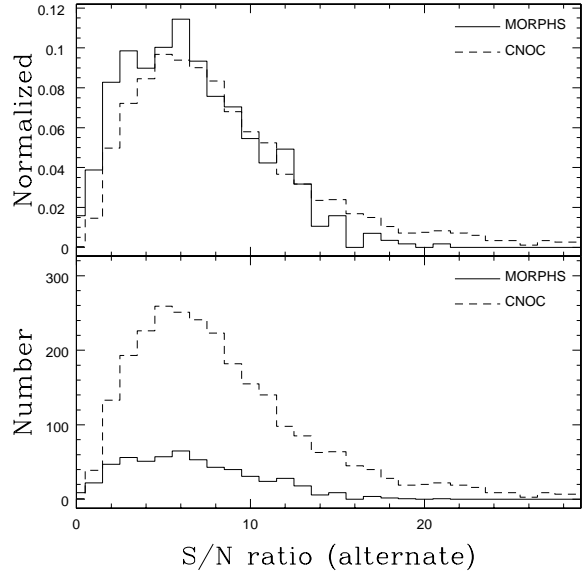}
\end{center}
\caption{We compare the S/N distribution for galaxies in the MORPHS sample with quality
indices less than 4{\it (solid line)}
to that in the CNOC1 sample {\it (dashed line)}.  In this case, S/N is measured as
the ratio of the mean flux per pixel in the range $4050<\lambda/\AA<4250$ to the {\it r.m.s.}
in the same range.  The {\it bottom panel} shows the number of galaxies in each bin;
in the {\it top panel} both samples are normalized to unit area.
\label{fig-SNcompare}}
\end{figure*}

In Figure \ref{fig-SNcompare}, we compare the S/N distribution of the CNOC1 sample to that
of the MORPHS sample (excluding from the latter galaxies that have a quality index of 4,
which are defined as those with S/N sufficient only for a redshift determination).  
In both cases, S/N is computed in the ``alternative'' manner 
described above.  Both samples have similar distributions, apart from a small excess of 
low S/N objects in the MORPHS sample\footnote{These S/N ratios are computed {\it per pixel}, 
and the CNOC1 spectra are sampled at 3.45 \AA\ per pixel, while many of the MORPHS spectra
are more coarsely sampled (for example, the COSMIC spectra are rebinned to 10 \AA\ per pixel).}.  
We can crudely assign an uncertainty to the index measurements made on the MORPHS sample
by correlating the errors on the CNOC1 data with our alternative S/N measurement.  From 
the S/N measured on the MORPHS data, we can then determine approximate index uncertainties.
Representing the uncertainties in \hd\ and \ow\ by $\Delta H\delta$ and $\Delta$[\ion{O}{2}],
respectively, we find $\Delta H\delta=9.2\times (S/N)^{-1.08}$ and $\Delta$[\ion{O}{2}]$=32.8\times (S/N)^{-1.22}$.

The uncertainty of a line index will be reduced
if the resolution is improved, and the continuum S/N ratio is fixed.
The MORPHS spectra were obtained using different telescopes and instruments, and are not
all of the same resolution; the best resolution spectra (obtained with the COSMIC
spectrograph) have a dispersion of
3.1 \AA\ per pixel, and are smoothed to the instrumental resolution of $\sim 8$ \AA, and
rebinned to 10 \AA\ per pixel.  Thus, the average FWHM of emission lines is comparable to that
of the CNOC1 spectra, $\sim 16.5$ \AA, as confirmed by the measurement of several [\ion{O}{2}] lines. 

\begin{figure*}
\begin{center}
\leavevmode \epsfysize=8cm \epsfbox{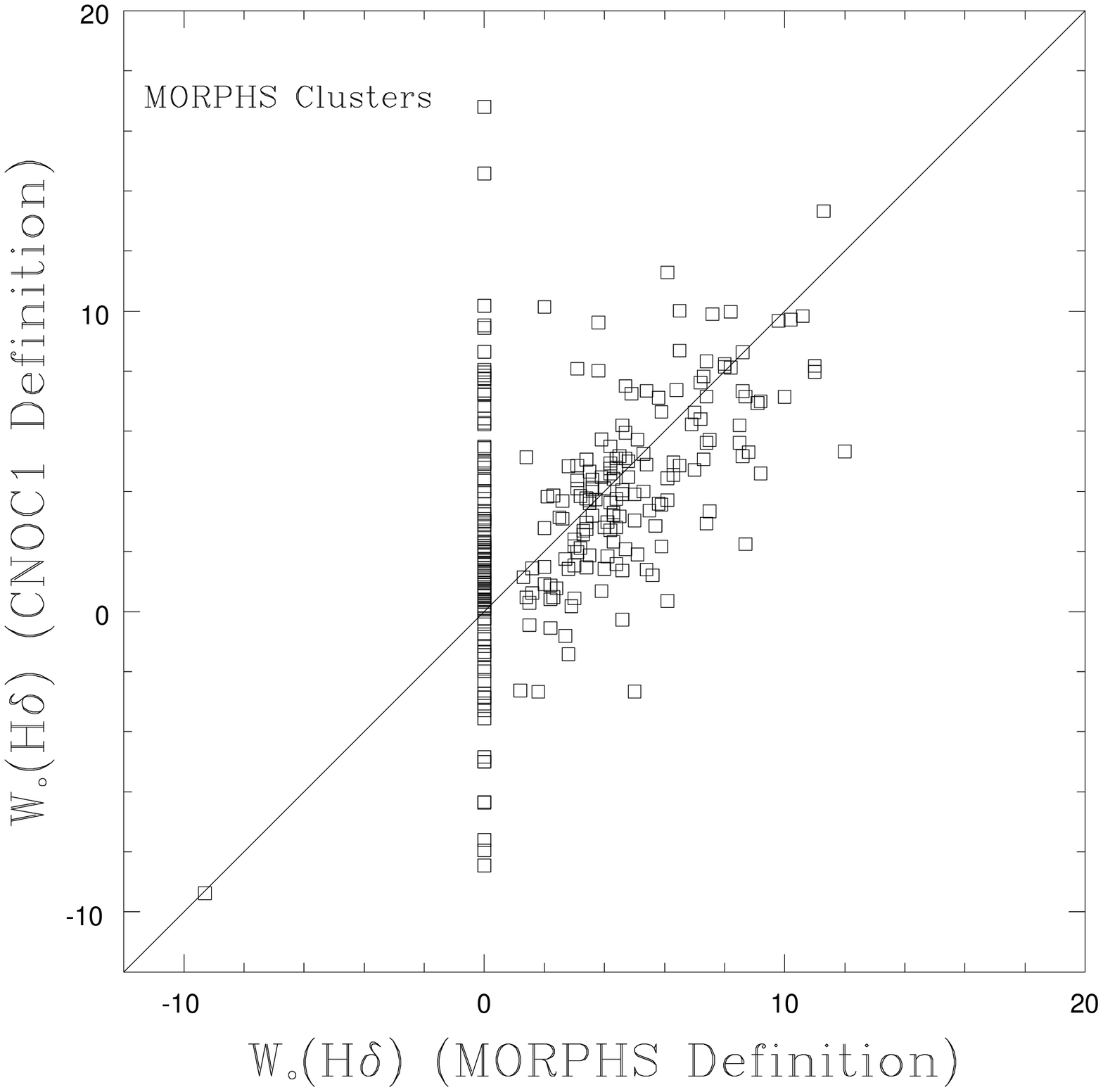}
\end{center}
\caption{Our measurements of \hd\ for galaxies in the MORPHS galaxy sample (on the y-axis)
are compared
with the equivalent widths published in Dressler et al. (1999, on the x-axis), excluding
galaxies with low quality indices ($q=4$) or flagged as uncertain in the catalogues.   Although the two indices
are defined in very different ways, they are correlated.  The {\it solid line} shows where the
two indices are equal; it is not a fit to the data.  The scatter in this figure reflects
the uncertainty in both indices, which arises from low S/N data and uncertainty in
continuum placement.
\label{fig-compareHd}}
\end{figure*}

A second property which must be tested is the equivalence of the CNOC1 line indices to those measured by
the MORPHS collaboration.  In Figure \ref{fig-compareHd} we show the values of our \hd\ index,
defined in \S\ref{sec-indices} and measured on the MORPHS spectra, compared with those of Dressler et al. (1999).  
We exclude galaxies with quality indices of 4, or flagged as uncertain.
The latter
were usually computed by interactively fitting the continuum and then fitting a Gaussian profile to the
absorption or emission lines.  Given the noisy nature of the data, and the generally
complex region of the spectrum under consideration, we find that such an interactive
approach is difficult to duplicate, and makes uncertainty
estimates difficult to determine reliably.  Nonetheless, the correlation between
our H$\delta$ index and that of Dressler et al. is fairly good.  In particular, we recover the
large \hd\ indices measured in many cases by Dressler et al., showing that our index is not
insensitive to spectra of this type.  The scatter in this figure
reflects the fact that both of these indices have considerable uncertainties.  

\subsubsection{Reanalysis of MORPHS Data}
\begin{figure*}
\begin{center}
\leavevmode \epsfysize=12cm \epsfbox{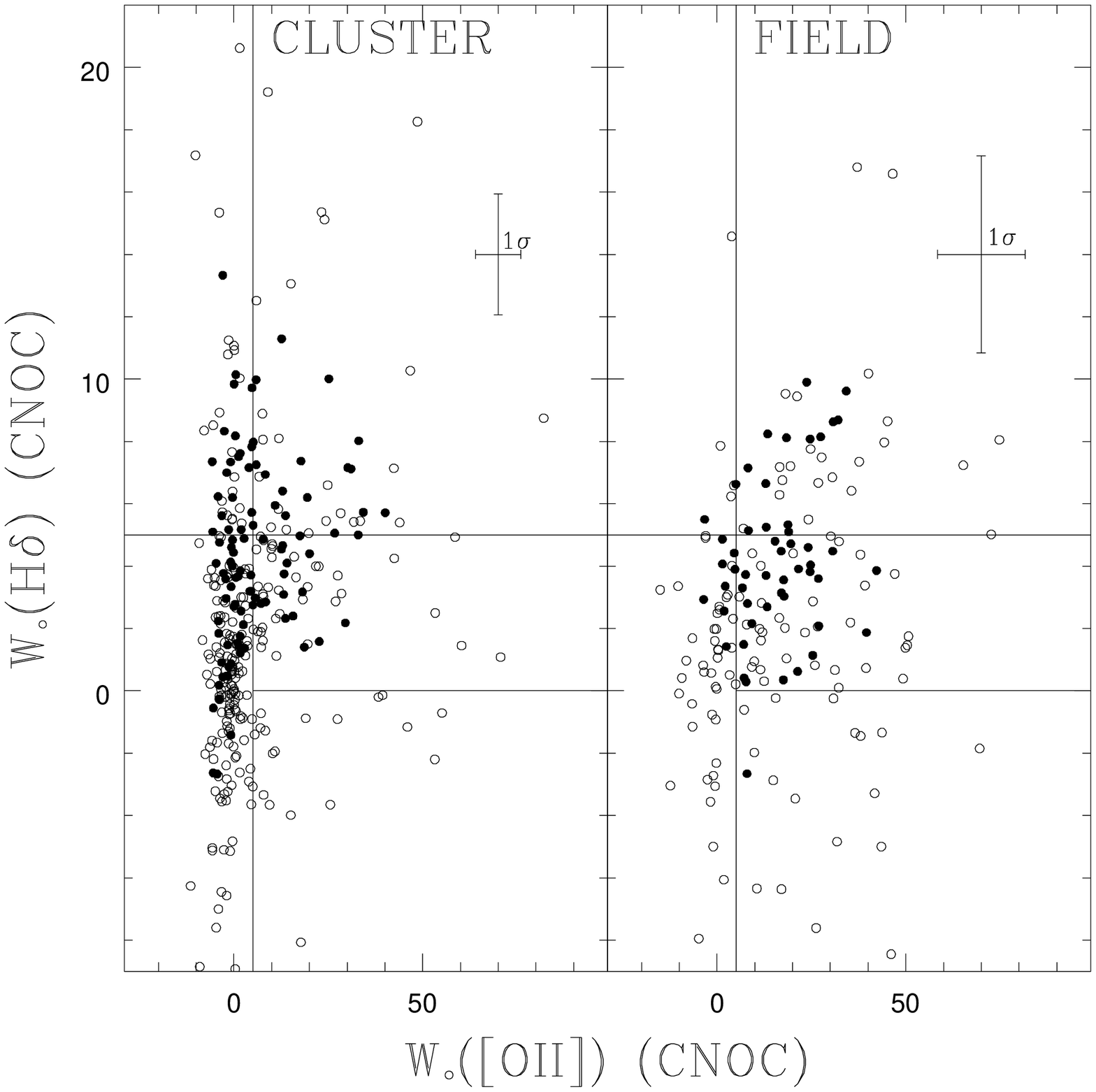}
\end{center}
\caption{Our measurements of \ow\ and \hd\ on the MORPHS 
cluster and field samples.  Data which are of quality index 4 have
been excluded.  {\it Open circles} are data for which either no \hd\ measurement is presented in the
MORPHS catalogue, or the measurement is flagged as uncertain. 
The sample error bars represent the mean, $1\sigma$ uncertainty, computed
assuming the CNOC1 correlation between S/N and index uncertainty holds for the MORPHS
sample.  
This figure may be compared with the CNOC1 data presented
in Figure \ref{fig-HdOII}, but note that the selection criteria are different
in that the MORPHS data are biased toward late type galaxies (which will generally have
stronger \ow\ and \hd\ indices).  There is a clear excess of K+A galaxies in the MORPHS clusters,
relative to their field sample.  Although the absolute fraction of K+A galaxies in this
sample is difficult to determine reliably, this excess suggests that the MORPHS clusters
may have different galaxy populations than the X--ray luminous clusters considered in the
present work.
\label{fig-HdOII_morphs}}
\end{figure*}

In Figure \ref{fig-HdOII_morphs} we show the MORPHS data in the \ow\--\hd\ plane, with
our region delineations from \S\ref{sec-defs-oiihd}, and
the measurements we have made using our index
definitions (\S\ref{sec-indices}).  We exclude data of quality index 4, 
and mark as open circles those galaxies which are either flagged as having 
uncertain \hd\ measurements, or for which no \hd\ measurement is listed in
the MORPHS catalogue (i.e., they are listed as either INDEF, or 0). 
Furthermore, no magnitude limit is imposed.  The sample error bars represent the mean error in
each index.  The field sample is somewhat biased to lower S/N galaxies and, hence, results in
larger mean index errors.

We will not attempt to evaluate the fractions of each type of galaxy (K+A, SF etc.) that
we have defined for galaxies in this plane, due to the difficulty in correcting for the morphologically
based selection (see Smail et al. 1997).  There is clearly, however, a strong presence of K+A galaxies in 
the cluster sample that is not observed in the field.  For most (20/24) of the galaxies with \hd$>10$ \AA,
Dressler et al. (1999) either did not measure \hd\, or the measurement was flagged as uncertain (open circles).
Often, this is due to poor sky subtraction near the line, which should result in large index uncertainties
that justify their removal from the sample (as we do for our maximal sample in \S\ref{sec-maxsamp}).  
However a large population of cluster K+A galaxies with \hd$<10$\AA\ persists.
This strong population is clearly not as abundant in the MORPHS field sample, in contrast with the
results found in our sample (Figure \ref{fig-HdOII}).  Due to the unusual selection criteria, however,
the absolute fraction of K+A galaxies in the two samples cannot be easily compared
in a quantitative way.  

\subsubsection{Possible Sources of Differences Between MORPHS and CNOC1}\label{sec-morphs-cnoc}
Given that the CNOC1 resolution, S/N distribution and line index definitions are comparable to those
of the MORPHS collaboration, what is the source for the difference in results?
We discuss three possibilities below.

{\bfseries Cluster Selection:} 
Given that galaxy clusters are a very heterogeneous class of objects, with differing
galaxy populations, some or all of the difference between the MORPHS and CNOC1 results may
be in cluster selection.  The CNOC1 clusters are among the most X--ray luminous 
at these redshifts, often containing cooling flows, and are mostly dynamically relaxed
(Lewis et al. 1999).  It is quite possible that, as a class, these clusters differ
from the more heterogeneously selected clusters in the MORPHS sample (though we confirm
K+A fractions of $\lesssim 5$\% for the two clusters in common with both samples).  
Furthermore, the
MORPHS clusters are generally at higher redshifts than the CNOC1 targets; all 
but four of the CNOC1 clusters lie at redshifts less than that of the lowest redshift
MORPHS cluster, A370.

{\bfseries Galaxy Selection: }
The selection of spectroscopic targets from the photometric sample differs in the
two surveys.
The MORPHS sample was primarily selected based on galaxy
morphology; Sd/Irregular galaxies are oversampled relative to earlier types.  
Since late--type galaxies were targeted preferentially, the detection of a large
population of starburst and/or post--starburst galaxies may not be surprising.
Dressler et al. (1999) attempt to correct for this by comparing the morphological
distribution of their spectroscopic sample with the photometric sample, 
but this correction is not completely reliable since the statistical
correction for late type field galaxies is uncertain (\cite{Smail}).
Alternatively, the primary selection criterion in the CNOC1 sample is apparent magnitude, with small corrections
for galaxy color and position. In particular, we note that
any color selection effect is quite small, and there does not
seem to be a significant bias to our results as a
function of galaxy spectral type at these apparent magnitudes.
As discussed in section \ref{sec-weights}, we used
weighting functions to correct our sample of line
indices to be representative of
the full photometric sample in the cluster fields.

\begin{figure*}
\begin{center}
\leavevmode \epsfysize=8cm \epsfbox{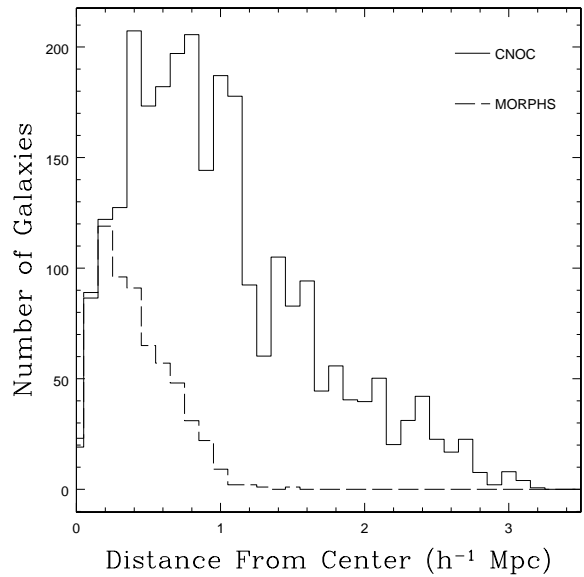}
\end{center}
\caption{The distribution of distances from cluster centers, for galaxies in the CNOC1 ({\it solid
line}) and MORPHS ({\it dashed line}) samples.  The CNOC1 distribution is weighted by
$W_{\rm spec}$ (see \S\ref{sec-weights}), but renormalized so that the total number of
galaxies represented in this figure corresponds to the number in the spectroscopic
sample.  On average, galaxies in the CNOC1 sample come from radii about 3 times larger
than galaxies in the MORPHS sample.
\label{fig-morphsrad}}
\end{figure*}

We also note that the CNOC1 survey includes galaxies at much larger
distances from the cluster center, as shown in Figure \ref{fig-morphsrad}.  
However, the lack of an observable gradient in the K+A fraction (Figure \ref{fig-HdOII_z})
makes it unlikely that this can have a strong effect.

{\bfseries Uncertainties:} No uncertainties are published
for the line indices presented in Dressler et al. (1999) or
Poggianti et al. (1999).  As we have shown in \S~\ref{sec-scatter}, the large 
uncertainties of our indices result in an overestimate of the K+A fraction;
since we believe the MORPHS index uncertainties to be comparable to ours,
the same holds true for their sample, and some correction for this effect must be made. 
Index measurements for galaxies with undetectable or hard to measure lines are not listed 
in Dressler et al. and, thus, the scatter in the measurements that do exist
does not accurately reflect the average uncertainty of the data.  

{\bfseries Galaxy Classifications: }Based on unpublished
recomputations of the Barbaro \& Poggianti (1996) models, using their new \hd\ definition, 
Poggianti et al. (1999) have chosen a threshold of \hd$>3$ \AA\ for their classification
of k+a/a+k galaxies. This is considerably
lower than we have adopted for our K+A definition, and clearly leads to the identification
of a larger population of ``unusual'' galaxies.  We have justified our limit not only by comparison
with the PEGASE models, but also by comparison with the local data of
Kennicutt et al. (1992a) and Kinney et al. (1996).  
From Figure \ref{fig-HdOII_local} alone, it is clear
that galaxies with \hd$>3$\AA\ and no detectable [\ion{O}{2}] emission are {\it not} exclusively K+As,
but include normal S0 and early spiral galaxies.  
Since the Dressler et al. \hd\ threshold is significantly lower than 
previously adopted by many other authors (e.g., \cite{DG83,FMB91,Z+96,A2390,1621}), 
we chose not to adopt their lower limit.  Adopting 3\AA\ instead of 5\AA\ as our lower limit for
the classification of K+A galaxies would increase the fraction
of galaxies that satisfy this definition\footnote{For reference,
7.6$\pm$1.1\% of cluster galaxies, and 9.1$\pm$1.6\% of field galaxies in our luminosity
limit sample have \hd$>3$\AA, and \ow$<5$\AA\ (uncorrected for scatter).}.  To some extent, of course, the choice of limiting
\hd\ is quite arbitrary and of little consequence as long as it is interpreted as such, and accounted
for when comparisons with other work is made.

\subsection{Summary}
With the exception of the MORPHS results (\cite{D+99}), the surveys discussed above
are not inconsistent out our principal results, which are:
(1) the fraction of K+A galaxies (as defined in \S\ref{sec-defs-oiihd}) in the field and X--ray luminous clusters at $z\approx 0.3$ 
is less than 5\%; (2) it has not been shown conclusively that the $z\approx0.3$ K+A
fraction is significantly larger than the fraction at $z\approx0.1$; and (3) there
is no strong evidence yet that K+A galaxies are more common in cluster environments than
in the field.  We suggest that the higher fractions of K+A galaxies
found by the MORPHS group may come from the very
different methodologies used, especially in the selection of the 
galaxy sample. We suggest that the simpler CNOC1 sample and weighting
functions are likely to produce a fair sampling of the cluster
populations.
In addition, there may be an actual difference in the
populations of our two differently selected cluster samples.
In that case, it remains to be seen whether
clusters at these redshifts are better typified by
the X-ray luminous CNOC1 sample, or the more heterogeneous MORPHS
sample.

\section{Discussion} \label{sec-discuss}
It was shown in Balogh et al. (1997) that the star formation rate (determined from
[\ion{O}{2}] emission lines) at all radii within
CNOC1 clusters is always less than that in the field galaxy population.  However, if a substantial
fraction of cluster galaxies have undergone massive but very short starbursts, 
as postulated for example
by Barger et al. (1996), the short duration  of these bursts ($\approx$100 Myr) would mean that only a
small fraction of cluster galaxies would be observed in the burst phase at a given epoch.

According to the results of the PEGASE models, following a short burst a galaxy
will appear as a PSB or PSF type for between 0.75 and 1 Gyr.  Thus, the starburst rate
can be determined from the abundance of these ``remnants''.  For  our maximal sample (\S\ref{sec-maxsamp}), the
fraction of these galaxies in the cluster is less than 6.2$\pm$0.7\%; this is an upper limit because it
does not include the correction for non-uniform scatter.  This implies that, in the last 2 Gyr, 
less than approximately 15$\pm2$\% of the galaxy population may have undergone a starburst.  
Using our best estimate for the scatter correction from Table \ref{tab-HDS}, the corrected fraction of cluster PSB and
PSF galaxies is 2.2$\pm0.9$\%, which reduces the estimated starburst population to only $\sim6$\% in the last 2 Gyr. 
This is much less than the 30\% estimated by Barger et al. (1996).  Part of this difference is due
to the factor of two difference between the abundance of HDS galaxies in our respective samples, as shown in
Figure \ref{fig-D4Hd} (significant at the $\sim 2 \sigma$ level).  Secondly, our fractions are
reduced due to the scatter correction, though Barger et al. include their uncertainties in their
models, which should account for this effect if their error bars are reliable.  Finally, Barger et al.
actually model the HDS galaxies (which comprise 4.6$\pm$1.8\% of our cluster sample, after scatter
correction) with post--starburst tracks; however, some of these have [\ion{O}{2}] emission (the A+em galaxies),
and cannot therefore be truly post--starburst.

Poggianti et al. (1999) have recently suggested that the A+em galaxies are dusty starbursts 
which may later evolve into K+A types;
the strong Balmer lines arise because the OB-stars responsible for the emission lines are heavily
obscured by dust, while the longer lived A stars have migrated out of their birthplaces and are
more widespread throughout the galaxy, so that their light dominates the spectrum.
In our cluster sample, we find  3.0$\pm0.9$\% of the galaxies have
A+em spectra, after scatter correction.  This is almost two times larger than the K+A
galaxy fraction and implies that the A+em phase must last at least twice as long as the K+A phase if the
two types are evolutionarily linked, and if {\it all} A+em galaxies evolve into K+A types.  
Thus, these dusty starbursts should either be fairly long--lived, with lifetimes of more than 1.5 Gyr,
or a large fraction of them never evolve into K+A types.  
The fraction of field galaxies classified A+em, $6.3\pm2.1$\%, is twice
as large as the cluster fraction.  Thus, the cluster environment does not appear
to be responsible for preferentially generating these galaxy types, whether or not they are starbursts.
It is also possible that, if this interpretation of A+em spectra is correct, some K+A galaxies
are just a more extreme example, in which the \oii\ line is completely obscured, as pointed out
by Poggianti et al.  In this case,
A+em and K+A galaxies are not evolutionary counterparts, but representative of the same, starburst
phenomenon. 

Assuming that K+A galaxies are the result of recently terminated star formation,
we can use the fractions of PSF and PSB galaxies in our sample to assess
the relative contribution of starbursts and truncated star formation to this scenario.
In general, PEGASE models of post--starburst galaxies spend roughly 
an equal amount of time in both PSB and PSF stages,
while truncated star formation leads only to PSF galaxies, lasting about 300 Myr after
the star formation activity ceases (see \S\ref{sec-defs-d4hd}).   Thus, if starbursts are the dominant mechanism for generating
HDS galaxies, there
should be (roughly) an equal number of PSB and PSF galaxies.  
As shown in \S\ref{sec-D4Hd}, the PSF fraction in both cluster and field environments may
be much higher ($\sim 5-10$ times) than the PSB fraction.   If this overabundance is real, it suggests that 
at least some of the galaxies may have had their star formation truncated without a burst,
as supported, for example, by recent simulations (Fujita et al. 1999).

The lack of a marked excess of K+A or A+em galaxies relative to the field may suggest that
the cluster environment does not actively 
truncate {\it or} induce star formation, but merely inhibits it from regenerating
once it has ceased. 
For example, in the galaxy formation model of Baugh, Cole \& Frenk (1996), 
after a merger of two spiral galaxies succeeds in producing
an elliptical galaxy, halo material may recollapse to form a new disk.  This reformation of 
a stable disk could conceivably be inhibited in the cluster environment, due to the presence of strong
tidal fields and galaxy harassment.  Furthermore, the halo of gas surrounding a spiral galaxy,
which may continually cool onto the disk, forming new stars, may be easily stripped 
away when a galaxy falls into a cluster (\cite{LTC}).  The conditions for this to happen are 
much less stringent than those for stripping all the gas out of the galactic disk.
In either of these cases, K+A
galaxies will generally evolve to become Passive within the cluster once they have 
exhausted their gas supply, whereas, in the
field, many will again become star--forming galaxies.  
Starbursts and K+A
phases are then interpreted as natural stages in galaxy evolution (i.e., galaxies
form stars in bursts, not continuously), and the cluster merely prevents 
a galaxy from recommencing its star formation once the activity has stopped of its
own accord.
We are currently analyzing dark matter simulations of clusters to link a galaxy's star formation history
to the time elapsed since it was accreted.  Preliminary analysis shows that a gradual reduction of SFR
following galaxy accretion adequately reproduces the radial gradient of star formation (Balogh et
al., in preparation).  

There are several caveats which may alter the conclusions described above, and there
are several ways in which future observations may address these issues; we list
these caveats below:

\begin{itemize}
\item[1.]Once statistical correction weights are included, our spectroscopic sample
is representative of a photometric sample complete to $M_r=-18.8+5\log h$.
When star formation is truncated in the more numerous, faint galaxy population,
they will fade and many will drop below this limit.  Indeed, the K+A galaxies 
in our sample appear to be of intrinsically low luminosity.  It is also known
that the less luminous galaxies evolve more significantly with redshift 
(\cite{CFRSVI}; \cite{Lin+97}), and thus contribute more to the Butcher--Oemler
effect.
Thus, it is possible
that there exists a large number of faint K+A galaxies which we are not detecting,
and these may show an environmental preference.

\item[2.] Whether or not an equal fraction of K+A galaxies in the cluster and field necessarily
implies an equal production rate of starbursts in both environments is somewhat
model--dependent, as it may only be a subset of the total population that is ever engaged
in starburst activity.  For example, the results of Ellis et al. (1997) and
Barger et al. (1998) suggest that 
most cluster ellipticals formed from a single
burst many Gyr ago, and have since evolved passively. 
If we assume that the Passive type galaxies play no part
in cluster evolution, then the fraction of ``active'' galaxies in clusters that
have a K+A spectrum is $6\pm3$\%, (corrected for scatter effects).
However, it is not yet clear how many of the Passive galaxies are safely excluded in this
manner; many of our Passive galaxies have spectra consistent with those of 
early type spirals and S0 galaxies.  Furthermore, it is
not clear what fraction of the field galaxies are also ``primordial'' (e.g., \cite{Kelson97}; \cite{Kodama98}).

\item[3.] 
The [OII] index is metallicity dependent, which affects star formation rates derived
from its measurement; however, this dependence is fairly weak (\cite{K92}) and unlikely to strongly
affect our results.  More important is the fact that this blue
feature is quite sensitive to dust obscuration, and many  of the PSB or PSF galaxies 
with \ow $\lesssim 5$\AA\ may
have significant, but dust--obscured, star formation (e.g., Poggianti et al. 1999). 
Infrared spectroscopy and imaging, and measurements of H$\alpha$ emission, 
will be helpful in constraining this effect.

\item[4.]The K+A population is known to be morphologically 
heterogeneous (\cite{C+98}; \cite{CRD}).  If high
resolution HST imaging reveals that K+A galaxies in clusters are morphologically
distinct from those in the field, this may support models in which the mechanisms
which are responsible for generating K+A galaxies are environment--dependent.

\item[5.]Kauffmann (1995) and Andreon \& Ettori (1999) have suggested that
the X--ray luminous clusters observed at moderate redshifts are not the
evolutionary predecessors of the less luminous, low redshift clusters to which
they are often compared.  In particular, the
CNOC1 sample is composed of the most X--ray luminous clusters in the universe
and, hence, may have unique or unusual galaxy populations; this may be responsible for some of the
discrepancy between our results and those of Dressler et al. (1999).  We note, however, that
Andreon \& Ettori find no dependence of the blue galaxy fraction on X--ray luminosity,
specifically.

\item[6.]Finally, the field sample drawn from the foreground and background of these rich
clusters may be unusual in some way, though results are generally consistent with the
field samples of other groups (\cite{Lin+97}).  Work is currently being undertaken to address this
issue using the CNOC2 field galaxy sample (Ellingson et al., in preparation). 
\end{itemize}

\section{Summary and Conclusions}\label{sec-conc}
We have presented a detailed analysis of the spectral characteristics of a large sample of
galaxies in X--ray luminous clusters between z=0.18 and z=0.55, and of an identically
selected field galaxy sample.  We focus on three spectral indices: D4000, which traces the
old stellar population; \hd, which indicates the presence of A--type stars and is
sensitive to star formation that took place up to 1 Gyr ago; and the \ow\ index,
which indicates current star formation activity.  We compare our data to the model predictions
of Fioc \& Rocca--Volmerange (1997) and Bruzual \& Charlot (1993), and draw the following
conclusions:
\begin{itemize}
\item The radial trends within the cluster sample are consistent with a continuous age sequence,
in the sense that the last episode of star formation occurred more recently for the outermost
galaxies than for the central galaxies. 

\item We define K+A galaxies as those with \ow$<5$\AA, and \hd$>5$\AA.
Our measured K+A galaxy fraction is $4.4\pm0.7$\% in the cluster and 
$1.9\pm0.6$\% in the field, but this is
only an upper limit due to the large index errors which tend to overestimate these numbers.
Attempting to correct for this effect, we find the true
fraction of K+A galaxies to be only 2.1$\pm$0.7\% in the cluster environment,
and 0.1$\pm$0.7\% in the field.  For our luminosity limited sample (galaxies brighter
than $M_r=-18.8+5\log{h}$), these corrected values are 1.5$\pm0.8$\% (cluster) and
1.2$\pm0.8$\% (field).
Our field results are consistent with the LCRS $z=0.1$ fraction of 0.30\% (\cite{Z+96}) at
the $\sim 1\sigma$ level.

\item  The fraction of cluster galaxies which are undergoing, or have recently undergone,
 a short burst of star formation is not significantly greater than the field fraction,
at any redshift or distance from the cluster center. 
From the fraction of PSB and PSF galaxies, we conclude that less than $\sim$10\% of
the galaxies in both the cluster and the field may have undergone short starbursts in the last 2 Gyr.

\item If all the A+em galaxies are dusty starbursts and progenitors of the K+A galaxies, then these
must be undergoing fairly long--lived episodes of star formation ($>1.5$ Gyr).  These galaxies are
two times more common in our field sample than in the cluster.

\item We find that PSF galaxies may outnumber PSB galaxies by a factor of 5-10, in both cluster
and field environments (though the samples are small).  Comparison with models suggests that truncated star 
formation {\it without} a short starburst phase may play a significant role in galaxy evolution.

\item More photometry of the CNOC1 fields is still
needed; particularly K--band to constrain the total stellar mass, and U--band
to independently measure star formation rates.  Infrared imaging and spectroscopy is needed to
determine the amount of dust--obscured star formation, and HST images are
required to subclassify the various galaxies types morphologically.

\end{itemize}
Keeping in mind the caveats listed at the end of \S\ref{sec-discuss},
we can discuss some of the implications of our results. The general trend
for higher redshift clusters to be bluer (the B--O effect) has not yet been
linked to a population of galaxies unique to the cluster environment.  The simplest
explanation consistent with the current data is that the B--O effect in clusters largely
reflects the increased level of star formation in the
field at larger redshifts.    Secondly, there is no evidence that recent ($<$1 Gyr) 
cluster--induced star formation
is responsible for driving the differential evolution between cluster and field
within the virial radius, since there is no tell--tale population of 
starburst or post--starburst
cluster galaxies that does not exist in the field.  
We speculate that the presence of bursting galaxies, and the possible increase of such galaxies with
redshift, reflects the general nature of star--forming galaxies, independent of 
environment. 
However, galaxies in dense environments are eventually {\em prevented} from bursting,
perhaps because their halo gas reservoir has been stripped.  This, coupled
with a population of ``primordial'' ellipticals and perhaps some truncation due to
ram pressure stripping of disk gas, could be responsible for the older stellar
populations which inhabit cluster environments.
 
\acknowledgments 
The data used in this paper form part of the CNOC1 study
of intermediate--redshift clusters.  We are grateful to all the consortium
members and to the CFHT staff for their contributions to this project.
MLB is supported by a Natural Sciences and Engineering 
Research Council of Canada (NSERC)
research grant to C. J. Pritchet and an NSERC postgraduate scholarship. 
MLB is grateful to J. F.
Navarro and A. Zabludoff for useful discussions.

\appendix
\section{A Description of the PEGASE Models}\label{sec-appendix}
In this Appendix we will discuss in more detail the PEGASE models shown in
Figure \ref{fig-models_local}, on which our galaxy classification system
(\S\ref{sec-defs-d4hd}) is partly based.

The large, open star in Figure \ref{fig-models_local} represents the model result
for a galaxy with a constant star formation rate, of any intensity.  It represents
the maximum H$\delta$ absorption one can expect in a normal galaxy; emission filling
will reduce its value by between roughly 1 and 3.5\AA\ (\cite{BP}).  This point lies 
blueward of our D4000 cut, within the SF region.  All of the galaxies in the SF
region are either spirals, irregulars, or ``starburst'' galaxies, and do not have \hd\
stronger than the model constant star--formation point (with the exception of NGC 3034, discussed in \S\ref{sec-defs-oiihd}).
Only if the star formation activity has begun recently (within the last 200 Myr, Barbaro \& Poggianti, 1997)
will H$\delta$ emission overwhelm the absorption and the galaxies will then occupy the short starburst (SSB)
region of this figure.  

The long--dashed line in Figure \ref{fig-models_local} traces the evolutionary path
of a galaxy with an exponentially decaying star formation rate; it evolves from ``birth''
at D4000$\approx1$, redward.  In this case, a decay time of $\tau=2$ Gyr is adopted, and
the galaxy is evolved to 11 Gyr, which we expect to be the maximum age of CNOC1 galaxies, in
reasonable cosmological models\footnote{Beyond an age of 11 Gyr, the models only increase in 
D4000 at a rate of about 0.05/Gyr.}. Most of the local spirals in the Kennicutt (1992b)
and Kinney et al. (1996) samples lie close to this curve.  Galaxies 
of this type with ages greater than about 5 Gyr will lie in the Passive region; clearly this includes not only
E and S0 type galaxies, but also early type spirals.  Note that the oldest models have \hd$<0$;
this reflects the nature of our specific index definition, which in this
case is telling more about features in the continuum, rather than the intrinsic H$\delta$
absorption (see \S\ref{sec-indices}).

The evolution of a galaxy produced by an initial burst of star formation
lasting 200 Myr is shown as the short--dashed line in Figure \ref{fig-models_local}.
After the burst ends, this spectrum always has stronger \hd\ than the exponentially
decaying star--formation model, until D4000$>1.7$.  It spends roughly 300 Myr in the
bHDS region and another 300 Myr in the rHDS region.  

The dotted line represents the evolution of
a galaxy which underwent constant star formation for 4 Gyr, followed by a 200 Myr starburst 
involving 30\% of its mass, after which all star formation was terminated.  This model track
only follows the evolution after the end of the starburst.  The model closely traces the
initial burst model, but does not reach as strong values of \hd.  The duration and strength
of the burst are quite arbitrary; longer bursts and bursts involving smaller amounts of
stellar mass result in weaker maximum \hd\ indices; this parameter space has been explored extensively
by Poggianti \& Barbaro (1996), though we warn that our \hd\ indices are defined 
differently and are not directly comparable with theirs.   

The rHDS galaxies can originate not only from a terminated SSB phase, but also 
from a galaxy in which long--term star
formation has recently ended (e.g., \cite{CS87,NBK}).  We model this latter case by terminating
star formation in a galaxy which has undergone constant star formation for 4 Gyr; this
is shown in Figure \ref{fig-models_local} as the solid line.  Once the star formation
is terminated, the galaxy quickly reddens, passing into the rHDS region within 100 Myr.
It then spends about 300 Myr in the rHDS region before \hd\ further weakens and
the spectrum becomes a Passive type.

\end{document}